\documentclass[10pt,letterpaper]{article}
\usepackage{opex3}

 \usepackage{subfigure}

\usepackage{amssymb}

\begin{document}

\title{Using ultra-short pulses to determine particle size and density distributions}

\author{Chris J. Lee$^{\ast}$, Peter J. M. van der Slot and Klaus. -J. Boller}

\address{Laser Physics \& Nonlinear Optics Group, Faculty of Science and Technology, University of Twente, P. O. Box 217, Enschede 7500AE, Netherlands}
\email{c.j.lee@tnw.utwente.nl}

\begin{abstract}
	We analyze the time dependent response of strongly scattering media (SSM) to ultra-short pulses of light.  A random walk technique is used to model the optical scattering of ultra-short pulses of light propagating through media with random shapes and various packing densities.  The pulse spreading was found to be strongly dependent on the average particle size, particle size distribution, and the packing fraction.  We also show that the intensity as a function of time-delay can be used to analyze the particle size distribution and packing fraction of an optically thick sample independently of the presence of absorption features.  Finally, we propose an all new way to measure the shape of ultra-short pulses that have propagated through a SSM.
\end{abstract}


\section{Introduction}\label{sec:Intro}
The pharmaceutical and food industries commonly processes particulates in solid-liquid and solid-gas mixtures.  However, real-time monitoring of the physical and chemical properties of these mixtures is a difficult task that may soon become a regulatory requirement under the FDA's PAT initiative \cite{FDA:PAT}.  Optical monitoring techniques are favorable because they are rapid and sensitive to many different production parameters.  However, common optical frequency domain techniques, such as Raman spectroscopy and near-infrared spectroscopy exhibit sensitivity to particle size, polymorphism, and chemical changes, which makes data interpretation problematic \cite{Blanco:2000,Pasikatan:2001,Taylor:1998,Patel:2000}.  For example, granulation, which is the aggregation of small particles into larger particles, presents a highly turbulent and energetic environment.  Apart from the desired particle size changes, it may also induce crystalline phase transitions, amorphism, and chemical changes.  These changes are usually rather difficult to distinguish from each other using frequency domain techniques \cite{Jorgensen:2004,Rantanen:2005,Taylor:2006}.  
   
The time-domain optical properties of strongly scattering media (SSM), such as imagining by time domain optical coherence tomography, has become the subject of intense study over the past 2 decades, spurred on by the availability of ultra-short pulses from modelocked lasers.  However, the common approaches to the analysis, which is an attempt to map the intensity of the scattered light as a function of time to geometric and physical properties of the scattering medium, can be complicated.  For instance, one can calculate the exact (Mie) scattering amplitude as a function of wavelength and viewing angle, assuming that scattering is dominated by a single type of particle with a known, regular shape (e.g., a sphere).  The approximate solution is then obtained by applying Monte Carlo techniques to a random ensemble of such scatterers, taking into account experimental details such as the viewing angle of the detector \cite{Hauger:1996,Mees:2001:1}.  Examples of this approach can be found throughout the literature including; aerosols\cite{Schaub:1991,Hoekstra:1994,Liu:2003,Mishchenko:2003}, biological tissue \cite{Schmidt:2000}, and latex spheres \cite{Hauger:1997}. This approach is widely employed in optical coherence tomography where the scatterers can be well approximated by spheres, and the experimental geometry can be carefully controlled \cite{Hauger:1997,Hauger:1996}.  However, in situations where the scatterers do not have a regular geometric shape, as found in crystalline powders, the initial calculation of the scattering amplitude becomes problematic.  This is because, one must numerically solve the Mie scattering equation for every shape of scatterers that are present, followed again by a Monte Carlo calculation with an ensemble weighted by the relative concentration of each shape.  Alternatively, it is possible to analytically or numerically solve the radiative transfer equations, for which there are numerous methodologies (see e.g., \cite{Vargas:1999}).  However, analytical solutions can only be obtained in very specific circumstances, such as when the diffusion approximation can be applied, while numerical solutions still depend on knowledge of the absolute shape of the scatterer.  These limitations make the application of the radiative transfer equations problematic, particularly in industrial processes, where the particle shape and the experimental geometry are not precisely known and difficult to control.

Less attention has been paid to random walk models, such as those developed by Bonner \emph{et al.}\cite{Bonner:1987}. However, in the case of pharmaceutical and food production, they are more clearly suitable since time-resolved particle size distributions, packing fractions, and the production of conglomerates in granulation processes are of interest, while imaging of the pharmaceutical formulation is less so.  In fact, knowledge of the particle shape cannot be employed in many pharmaceutical processes since they often involve multiple components with very different shapes which may change during the manufacturing process.  Examples of such processes include milling, mixing, and granulation.  Thus, a fast and effective analysis technique that does not depend on knowing the shape of the scatterers is desirable.  This tool could then be used to separate geometrical parameters such as particle size distribution and packing fraction from other parameters, such as crystalline phase transitions, and chemical changes. 

In this paper we present a simplified random walk, ray tracing approach to analyze the scattering dynamics of random media in response to ultra-short pulses of light.  The simulated medium is characterized solely by its particle size distribution, packing fraction, and light propagation \emph{within} the particles.  The dependence on particle size range, distribution, packing fraction (the density of scatterers), and absorption are examined.  We show that by measuring the time-resolved intensity of the backscattered radiation, it should be possible to separate out the influence of packing fraction, particle size distribution, and particle size range from each other.  The inclusion of light propagation within the particles allows us to examine both the influence of a wavelength dependent optical path length, and absorption within the particles on the temporal shape of the output pulse (aside from the obvious loss of photons).

\section{Approach}\label{sec:Approach}
The approach adopted here is similar to that of Bonner \emph{et al.} \cite{Bonner:1987}, where scattering is treated as a ray tracing problem, with the randomization of the ray trajectory on a continuum of lengths and angles \cite{Gandjbakhche:1993,Bonner:1987}.   Each ray is referred to as a photon to make references to optical dispersion, absorption, and time dependence less convoluted.  The essence of our approach is that we extend \cite{Bonner:1987} to include several significant factors that are relevant to industrial practices; the influence of particle size distribution, packing fraction, and the temporal shape of the backscattered light pulses.  In addition, we include optical dispersion and influence of absorption features on the dispersion.  Where, in this paper, dispersion refers to the influence of refractive index, rather than pulse stretching due to multiple paths.  We do not include photon loss due to absorption, since, in the near infrared, where modelocked lasers are common, absorption is usually due to overtone modes and is very small.

The particle system is modeled by a fractional density, $\rho_p$, and a particle size distribution, $d_p$, where the size of an individual particle is given by the diameter of the bounding sphere.  The particles are also considered to have a refractive index, $n_p(\lambda)$, where $\lambda$ is the wavelength of the illuminating light, in order to associate an optical size with the particles.  The gaps between the particles are considered to be dispersionless with a refractive index of unity, the gap density is given by $\rho_{g} = 1-\rho_p$, and the gap size distribution is related to the particle size distribution by $d_{g} = [\rho_{g}d_p^3/(1-\rho_{g})]^{1/3}$.  Ray tracing is performed over a limited volume defined by $0 \leq z \leq 2$~mm, $-1.0 < x, y < 1.0$~cm.  The photons (rays) initially enter a particle at the $z = 0$ surface within a given time interval and with a temporal distribution that corresponds to the temporal profile of the injected ultra-short pulse.  This is functionally equivalent to assuming a flat front surface, however, our results do not change when this condition is relaxed and it corresponds well to certain physical systems where a powdered system is compacted at high pressure \cite{Pederson:1994, Sun:2006}.  Thereafter the photons are free to exit the volume in any direction, however, only those photons that exit through the $z = 0$ surface are considered to be detected.  The sample depth can be an important consideration for scattering studies.  Even with the largest particles (325~$\mu$m), only a small amount of light ever exits the rear face (1--5\%).  This can be understood by looking at the maximum probability of traversing the sample volume in the minimum number scattering steps ($1/2^6$). For the rear face to affect the backscattering signal, the light must return from the rear interface, which also has a maximum probability of $1/2^6$. Indeed, we found that the results presented in Fig. \ref{fig:Pulsedynamics} are independent of sample thickness, provided the amount of transmitted light is small.

Instead of defining each particle with an independent shape, a random free path that depends on the particle size distribution is generated for each photon and its direction is randomized to model scattering at the particle interface.  To obtain the temporal shape of the backscattered light pulses, those photons that exit the $z$=0 face of the sample are then used to generate a histogram of photon counts as a function of time. In most simulations performed by others, an exponential distribution, corresponding to processes that have a constant chance of occurring per unit time, is used \cite{Vellekoop:2005}, which is only true for very finely spaced scatterers.  In our simulations we assume that; the particles are much bigger than the wavelength of light, and are of comparable size or bigger than the length of the injected pulse, which is a situation relevant to processing powders in an industrial situation.  In this case, the probability of scattering at a particle-gap interface is nearly unity, while the probability of scattering within the particles is much smaller, and indeed negligible for crystalline material with very few inclusions.  Hence, we can equate the random free path distribution with the particle size distribution, which is sufficient to explore the influence of dispersion and absorption, and to qualitatively observe the influence of particle size and packing fraction on the temporal shape of the scattered light.

\begin{figure}
	\centering
	\subfigure[]{\includegraphics[width=5cm]{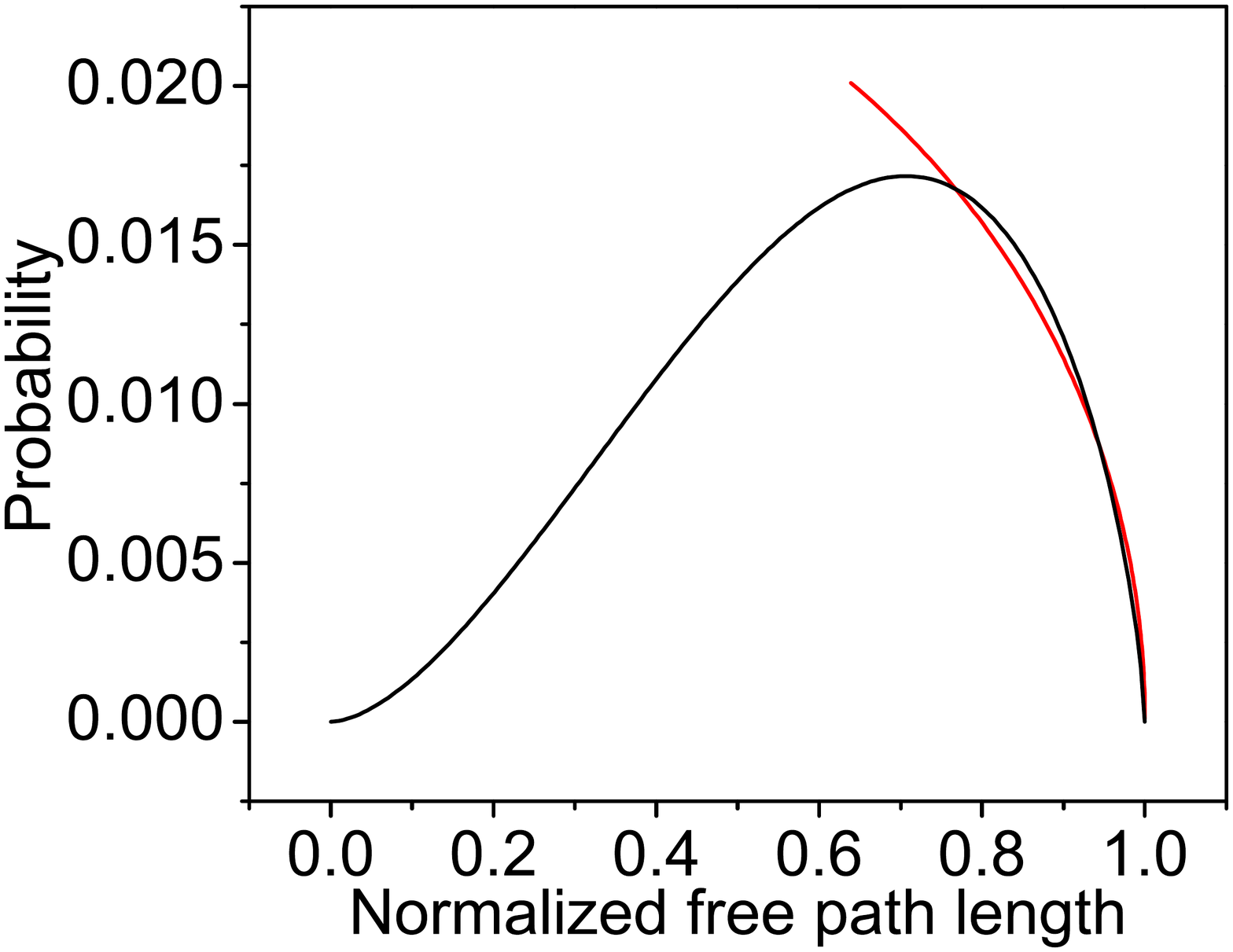}}
	\subfigure[]{\includegraphics[width=5cm]{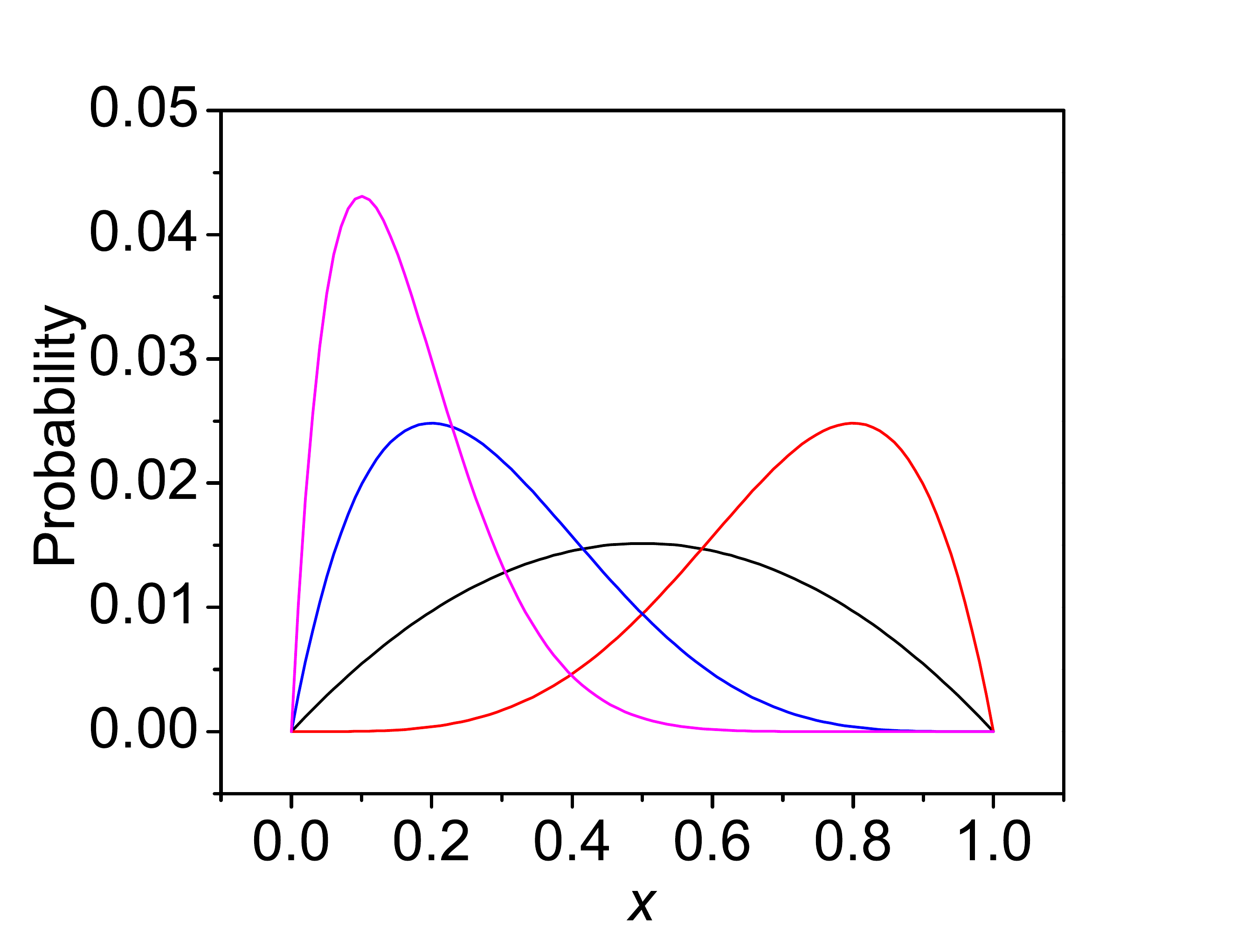}}
	\caption{The probability of a ray traversing a sphere along a particular chord length (a).  The calculated data from Snell's law is given by the black line, while the red line is equation~\ref{eqn:beta} with $\alpha=2.7$ and $\beta=1.7$.  Subfigure (b): equation~\ref{eqn:beta} for $\alpha = 2$, $\beta = 2$ (black), $\alpha = 5$, $\beta = 2$ (red), $\alpha = 2$, $\beta = 10$ (magenta), and $\alpha = 2$, $\beta = 5$ (blue).}
	\label{fig:probDistribution}
\end{figure}

So far, the approach described here assumes that, once a photon has entered a particle, there is a well-defined path length, given by the bounding sphere of the particle, between scattering events. However, this is not very likely, as demonstrated in Fig. \ref{fig:probDistribution}(a), because, even in spherical particles of diameter $d$, the path $x$ taken through it is not always the longest available ($d$).  Instead, there is a distribution of path lengths, $x$, of which only the longest is $d$. In order to take this effect into account we introduce the following free path distribution for photon scattering within a particle,  
\begin{equation}\label{eqn:beta}	
f(x;\alpha,\beta) = \frac{1}{\mathbf{B}(\alpha,\beta)}x^{\alpha-1}(1-x)^{\beta-1}
\end{equation}
The path length distribution, $f(x,\alpha,\beta)$, is only valid on the interval $0 \leq x \leq 1$ (see Fig. \ref{fig:probDistribution}(b)) , which is a normalized particle dimension, and the probability distribution, $f$ is normalized to unity by
\begin{equation}\label{eqn:betaNormal}
	\mathbf{B}(\alpha,\beta) = \int_0^1{t^{\alpha-1}(1-t)^{\beta-1}dt}
\end{equation}
The parameters $\alpha$ and $\beta$ can be set to shape the distribution in order to roughly refer to the particle shape, e.g., as for needle-like particles ($\beta > 2$)  or spherical (see Fig. \ref{fig:probDistribution}(a)) and cubic shape ($\alpha > 2$).  Other distributions are also possible, however, independent of such choice, due to the normalization, the distribution reflects the most essential physical property that scattering occurs at the particle-gap interface with unity probability, and that the path through a particle is not always the longest available. We have investigated the influence of various free path distribution on the temporal shape of the scattered light in section \ref{ssec:fpd}, by varying $\alpha$ and $\beta$. However, other types of normalized distribution functions may also be useful to address certain types of particles.

The time dependent intensity of a single, incident laser pulse is simulated by a photon number, which was generally set 10$^6$ photons, distributed evenly over a spot of 1~mm$^2$ area and with starting times that follow a Gaussian distribution with a 50~fs width (FWHM).  For each simulation, the light was considered to have only a single frequency and all rays enter the sample volume parallel to the $z$ axis.
 
The described ray tracing technique has the advantage of being computationally simple, which allows for it to be quickly generalized to many different scattering geometries.  Only the calculation of some simple trigonometric relationships and the generation of three random numbers per photon per interface is required.  At each interface, the time and position of the photons remaining in the medium are updated.  Thus, the program progresses in steps of constant scattering events, rather than in steps of constant time and gets increasingly faster as the simulation progresses and photons leave the sample volume.  After all the photons have left the sample, the exit times of those photons that left the $z = 0$ face of the sample are used to create a histogram of photon number as a function of time.

The technique also has some disadvantages.  For instance, the Lorenz-Mie scattering theory predicts that certain shapes and sizes will have enhanced scattering cross sections for particular wavelengths, due to interference effects, particularly in monodisperse media with scatters of uniform shape.  Since no particle shape is defined in our approach, these resonances will not be present in our simulations and the results cannot be used reconstruct the shape of the scatterers.  However, we believe that this lack of detail is of minor importance in the situation of interest here, where particles are typically polydisperse and of nonuniform shape.

\section{Results}\label{sec:results}
Using the model described in section \ref{sec:Approach}, the change in shape of an optical pulse by backscattering from the sample and its frequency dependence are analyzed under several different conditions.  In subsection~\ref{ssec:no_dispersion} the pulse stretching due to a non-dispersive, non-absorbing medium is examined under the condition that the free path distribution is the same as the particle size distribution.  Although the results in this section are an oversimplified version of what can be expected in experiments, it clearly illustrates where the dependence of packing fraction and particle size will manifest.  Note that we explore scattering for packing fractions between 0.125 and 0.9.  The upper limit that is typically associated with random packing is readily exceeded in pharmaceutical process such as tableting, where high pressure is used to compact the scatters to near 100\% of the component's natural density \cite{Pederson:1994, Sun:2006}.  In subsection~\ref{ssec:dispersion} the influence of optical dispersion is added, demonstrating that, in our simulations, the effect of dispersion is negligible compared to the influence of particle size.  Subsection~\ref{ssec:dispersion} examines the influence of absorption on the temporal shape of the scattered light.  The influence of the free path distribution within particles is examined in subsection~\ref{ssec:fpd}.  Unless otherwise noted, the medium is considered to consist of scatters, the size of which are uniformly distributed over a specified size range (flat-top size distribution), as one might expect from sieving.
 
\subsection{Non-dispersive, non-absorbing medium}\label{ssec:no_dispersion}
\begin{figure}
	\centering
	\includegraphics[width=7cm]{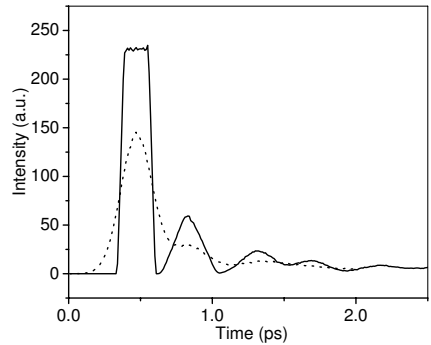}
	\caption{Exit pulse from a random media with a packing fraction of 0.5.  The Gaussian input pulse is centered around 0~ps and is 50~fs in duration (FWHM).  The solid line represents the response from a system of particles, uniformly distributed in size from 85 to 135~$\mu$m.  The dotted line is the response from a system of particles with a Gaussian distribution centered on 110~$\mu$m, with the half-width of the distribution at 85 and 135~$\mu$m.}
	\label{fig:Pulsedynamics}
\end{figure}
In Fig.~\ref{fig:Pulsedynamics}, two typical calculated output pulse shapes are shown.  The Gaussian input pulse is centered around $t$~=~0 and is 50~fs (FWHM) in duration (not shown) and is considered to have a wavelength of 800~nm.  The solid line shows the scattering from a system that consists of particles with sizes uniformly distributed between 85 and 135~$\mu$m.  In contrast, the dotted line shows the scattering from a system of particles with a Gaussian distribution (centered on 110~$\mu$m, with a FWHM of 50~$\mu$m).  As can be seen, the pulse shape of the incident fs pulse is stretched and delayed into the ps range by the scattering, with the main peak broadening and the development of several subsequent peaks.  The time delay between the input pulse and exit pulse is due to the time between entering the sample and the first possible scattering event, while the following peaks represent contributions from photons that return from deeper within the particulate system.  It is also apparent that the output pulse shape is the same as that of the particle size distribution.  This is because the output pulse is essentially a convolution between a delta function and the particle size distribution.  Although this is a direct result of approximating the free path distribution with the particle size distribution, the output pulse will always be a convolution of the input pulse with the free path distribution.  Thus, as long as the free path distribution can be linked to the particle size distribution, the particle size distribution can be obtained from the shape of the main pulse from the backscattered light.  

\begin{figure}
	\centering
		\subfigure[]{\includegraphics[width=5cm]{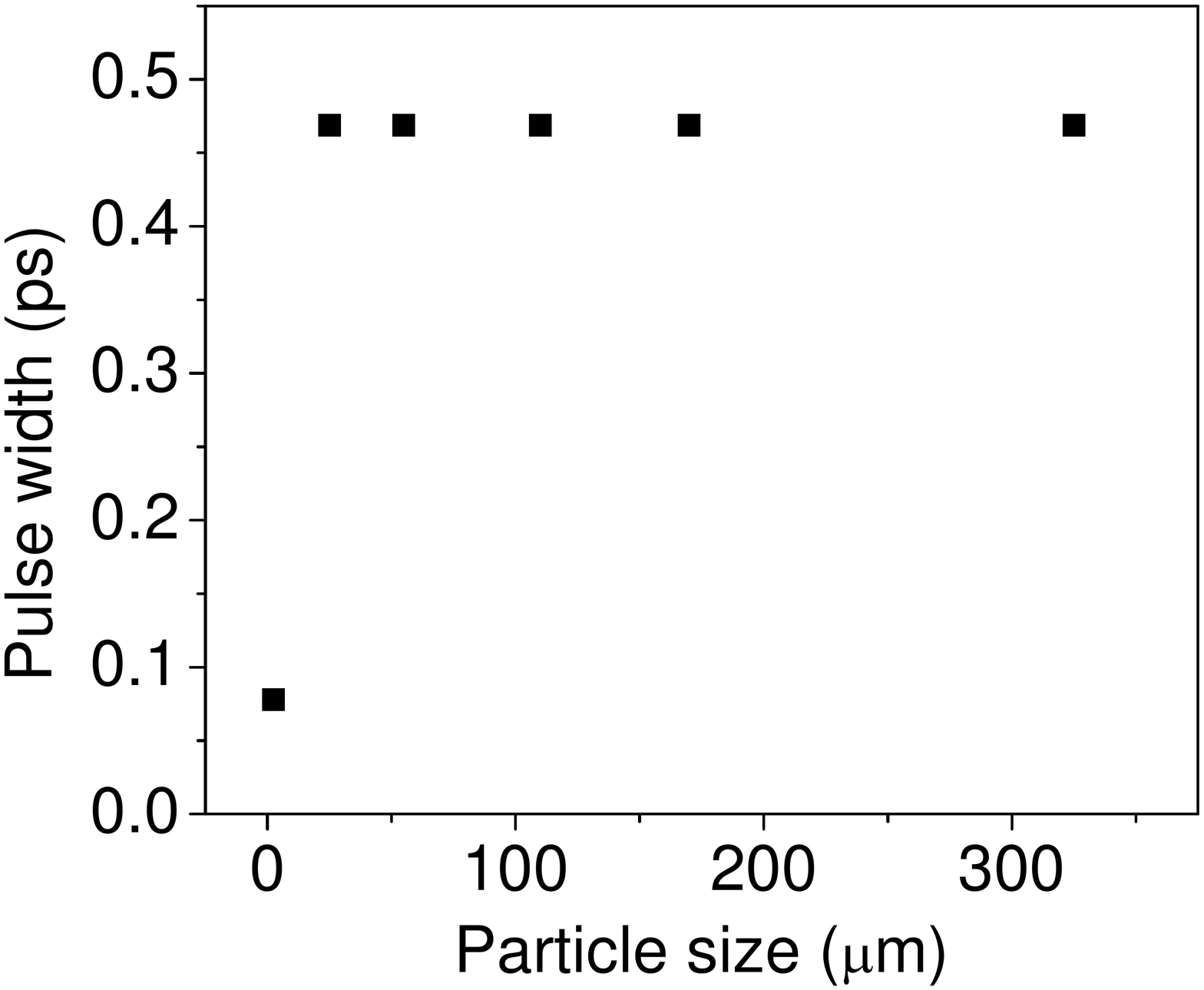}}
		\subfigure[]{\includegraphics[width=5cm]{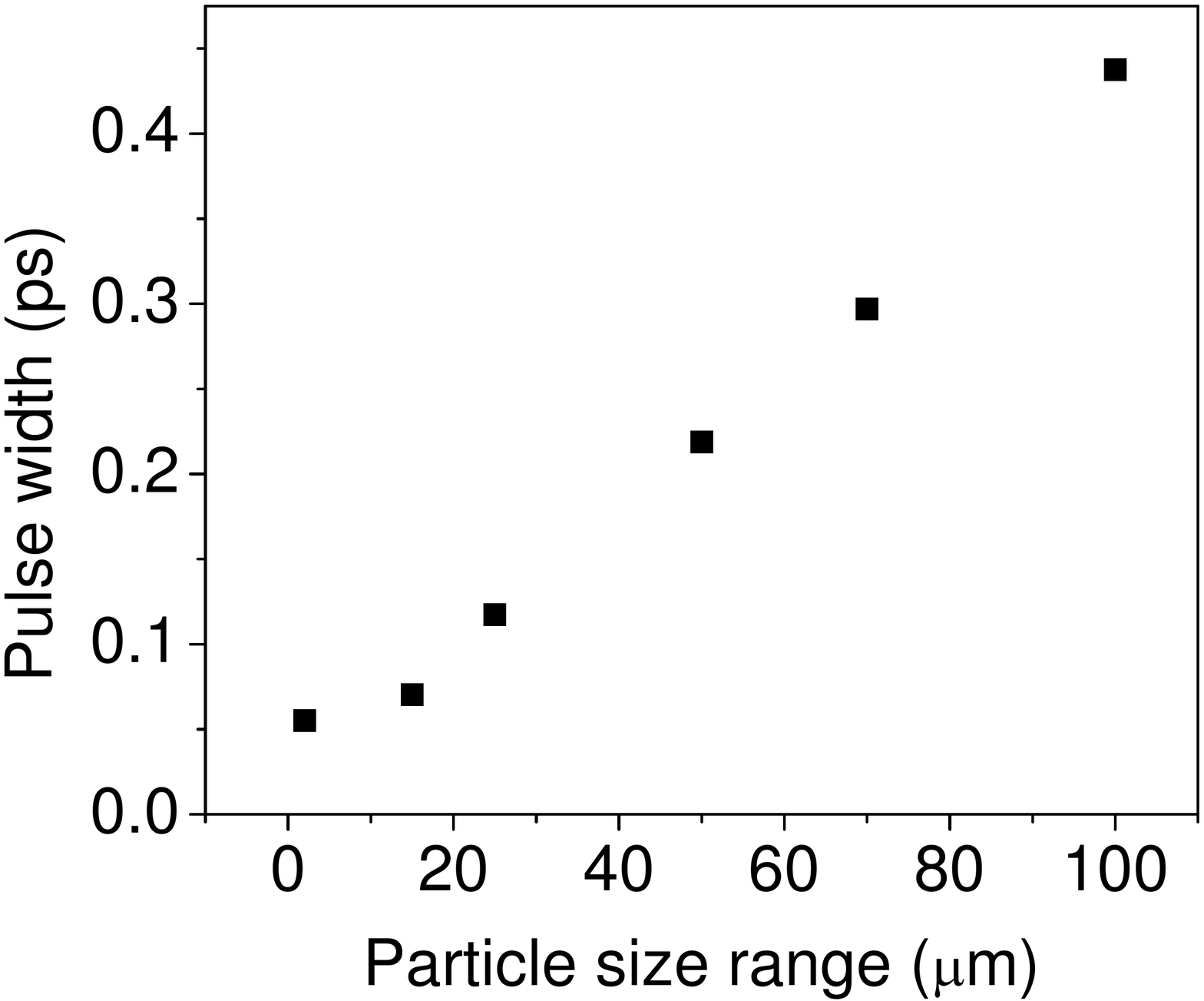}}
		\caption{Pulse spread as a function of particle size and size range.  For all simulations, the packing fraction was set to 0.5.  Pulse spread as a function of particle size (a).  The particles were uniformly distributed over a 2~$\mu$m range centered on the marker.  Pulse spread as a function of size range (b).  The particle sizes were uniformly distributed about 110~$\mu$m. }
	\label{fig:pulseDispersion}
\end{figure}

The width of the main return pulse depends only on the width of the particle size distribution, which can be seen in Fig.~\ref{fig:pulseDispersion}.  Here, the width of the main pulse is plotted as a function of particle size (a), and as a function of the width of the size distribution (b), for a constant packing fraction of 0.5.  Particle sizes from 2.5 through to 325~$\mu$m were assumed, each with a size range of 2~$\mu$m.  It can be seen that the pulse width of the main pulse is independent of the particle size for all sizes, except the smallest.  Here, the reflected signal and the signal from internally scattered light could not be distinguished at the temporal resolution employed in the simulations. The pulse spreading was also found to be independent of packing fraction, which means that, in principle, a measurement of the pulse spreading of the first return pulse is enough to determine the particle size range, independently of packing density and the median particle size.

\begin{figure}
	\centering
		\subfigure[]{\includegraphics[width=5cm]{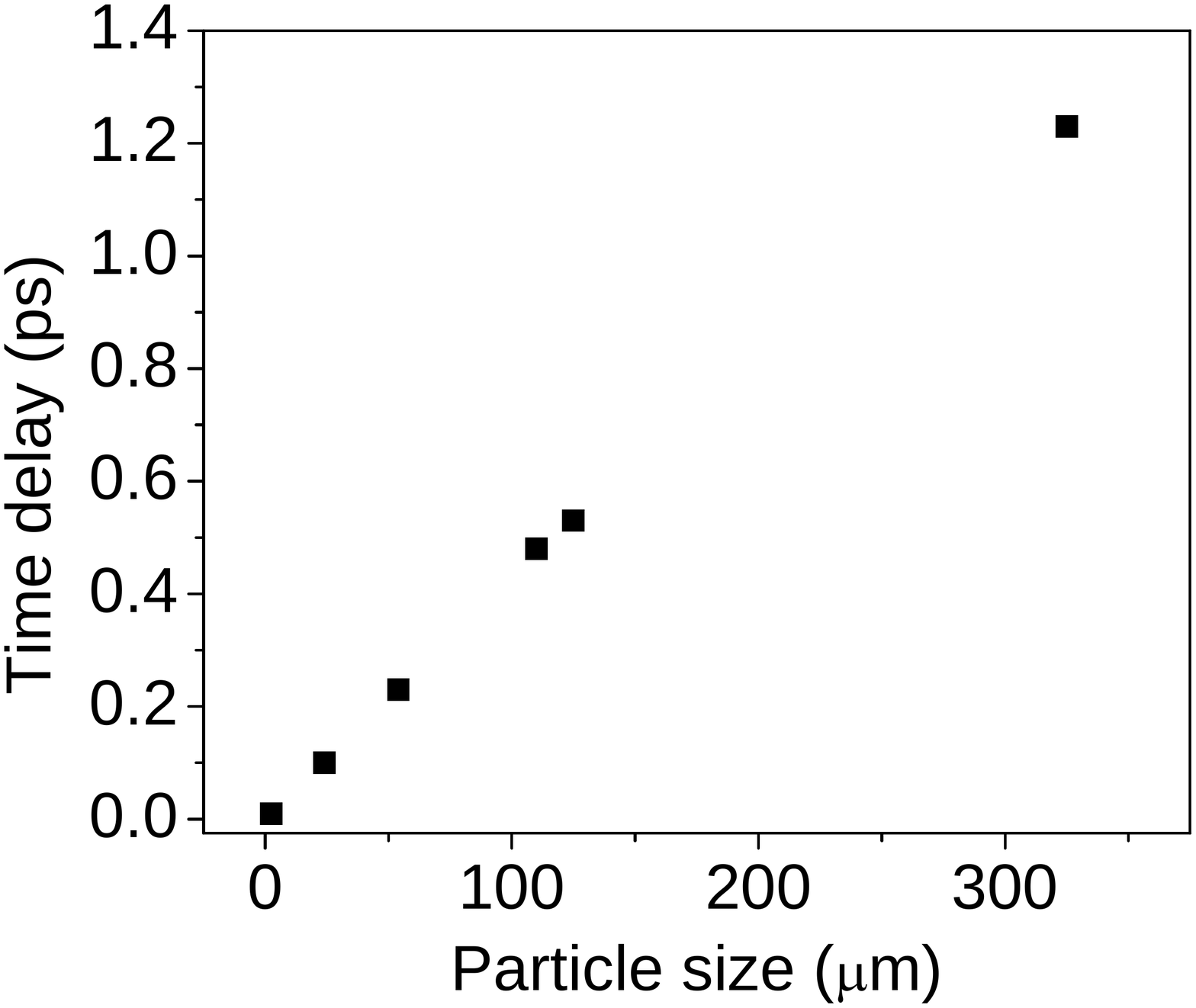}}
		\subfigure[]{\includegraphics[width=5cm]{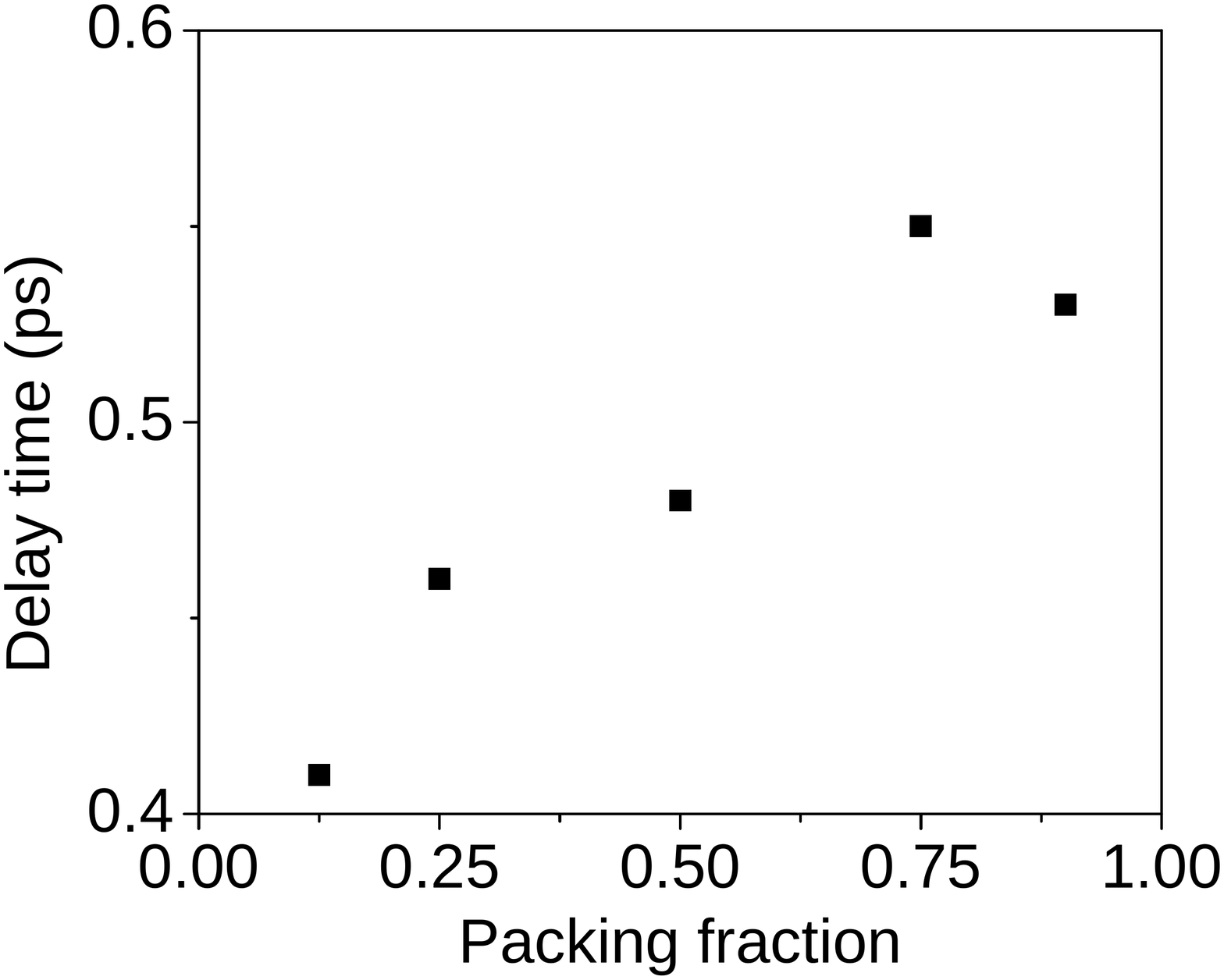}}
	\caption{The delay time between input pulse and main return pulse as a function of particle size (a).  The packing fraction was 0.5.  The delay time between input pulse and main return pulse as a function of packing fraction (b).  The particle size was 110~$\mu$m.  For both simulations the input pulse width was 50~fs.}
	\label{fig:DelayTime}
\end{figure}

The delay time between the input and main pulse increases linearly with particle size (see Fig.~\ref{fig:DelayTime}(a)), which is expected.  This measurement is essentially a time of flight measurement between the arrival of the photon at the scattering system and the average number of scattering events required to return a photon.  Since the number of scattering events per unit distance goes up with decreasing particle size, it is expected that smaller particles should return a pulse sooner than larger particles.  It was also expected that the delay time should depend on packing fraction, though not linearly because the mean fee path in the voids between particles scales as $(1/\rho)^{1/3}$.  Figure~\ref{fig:DelayTime}(b) clearly shows that the delay time \emph{increases} with increasing packing fraction.  This is due to the higher optical density of the medium compared to the voids.  The gaps must be approximately one third larger than the material particles before the photons experience similar travel times between scattering events.  Therefore a more densely packed medium will have a longer time delay between scattering events since the photons spend most of their time in the more optically dense material.  The delay time apparently decreases slightly for very high packing fraction.  This may indicate that the packing fraction behavior is more complicated than indicated and this will be the subject of future simulations.

\begin{figure}
	\centering
		\subfigure[]{\includegraphics[width=5cm]{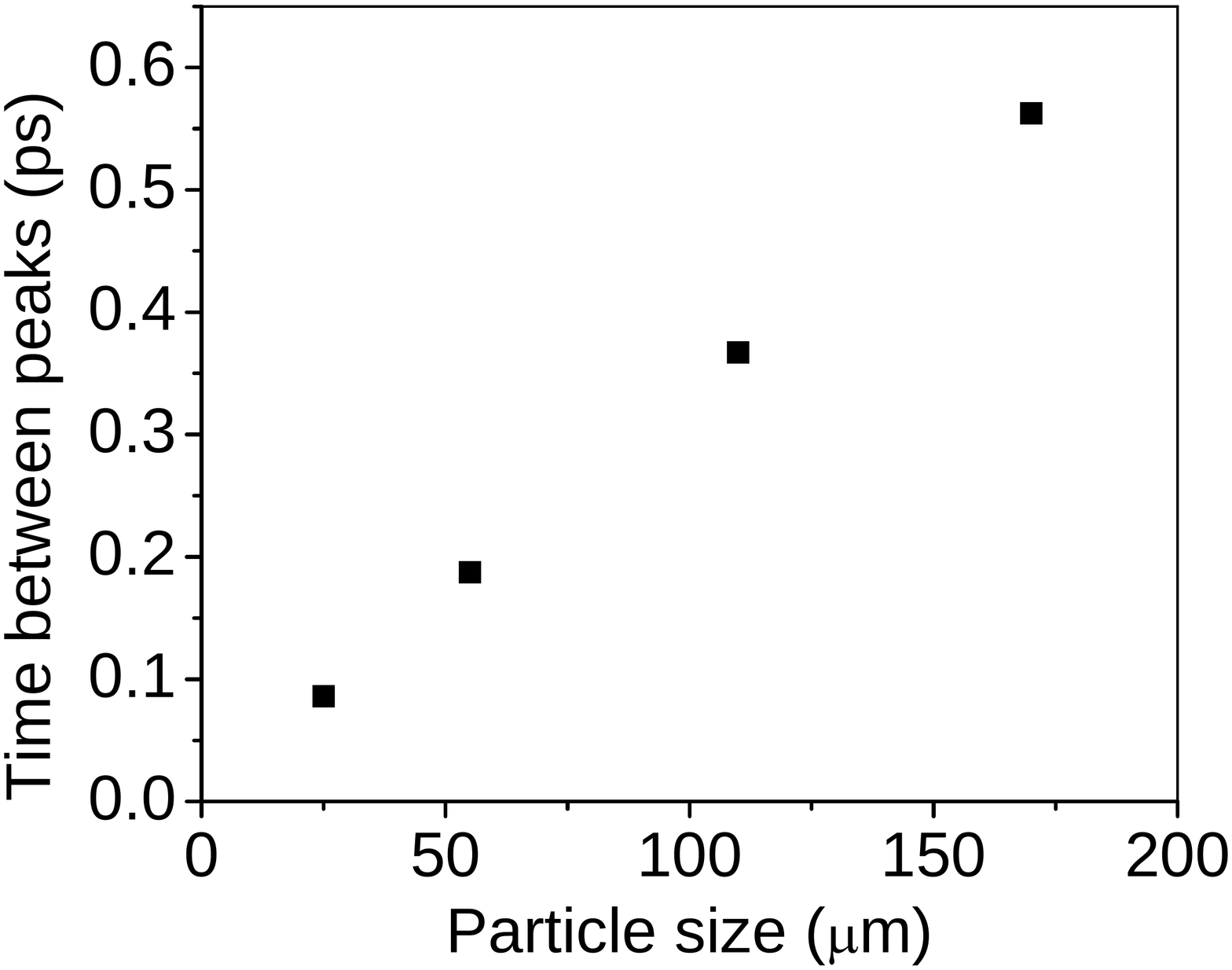}}
		\subfigure[]{\includegraphics[width=5cm]{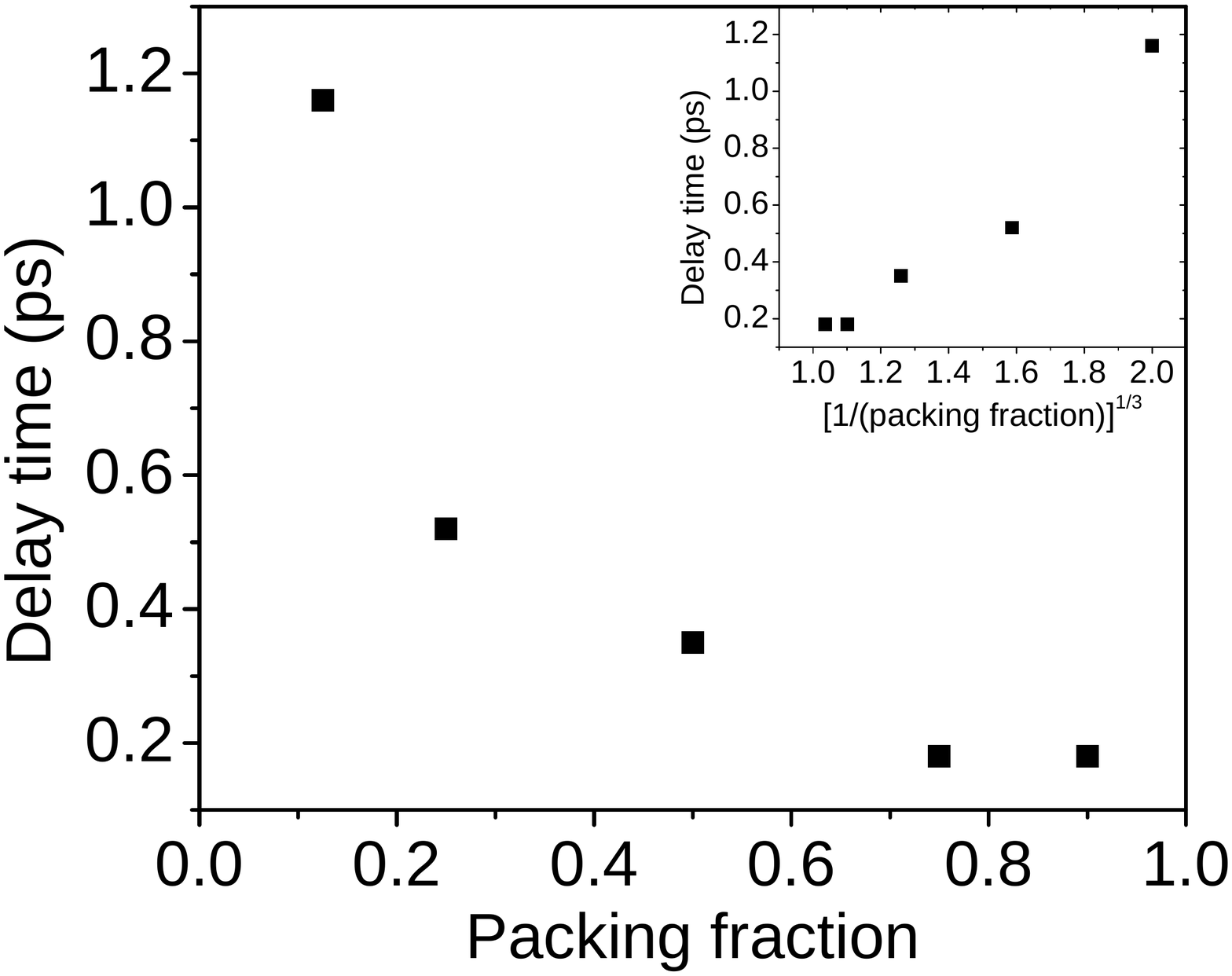}}
	\caption{Delay time between input pulse and main return pulse as a function of particle size at a constant packing fraction (0.5) (a).  Delay time between input pulse and main return pulse as a function of packing fraction at constant particle size (110~$\mu$m) (b).  The inset shows the dependence on the inverse cube root of packing fraction, which is expected to be linear.  For both simulations the input pulse width was 50~fs.}
	\label{fig:DelayBetweenPeaks}
\end{figure}

As shown in Fig.~\ref{fig:Pulsedynamics}, the output pulse can consist of a main peak with several subsequent peaks.  We find that the delay between the first two peaks is dependent on both the particle size and the packing fraction (see Fig.~\ref{fig:DelayBetweenPeaks}).  The delay between the first two peaks is linearly dependent on particle size for large particles.  Again this is because the measurement is a time of flight measurement. 

Likewise the delay between peaks varies with packing fraction as well.  In this case the delay time decreases with increasing packing fraction.  This can also be explained by time of flight considerations, as a reduced packing fraction indicates that a photon must travel quite far before encountering an interface to scatter from.  In this case the dependence is clearly not linear, but rather it varies as $(1/\rho)^{1/3}$, which is expected.

\subsection{Dispersive, non-absorbing medium}\label{ssec:dispersion}
The simple simulations above show that backscattering from ultra-short pulses has some promise for rapidly characterizing SSM.  However, these simulations did not include material dispersion, which may become necessary when one considers that ultra-short pulses cover a finite spectral range.  

To consider the effect of dispersion the same simulation is used, however, the optical path length of each ray within particles was varied over a certain range in a sequence of simulations, to express the wavelength dependent refractive index of the particles.  As an example, we used the refractive index for crown glass \cite{SchottCat:2006}.  The wavelength of the light is centered on 800~nm with a width of 120 nm, divided into 256 spectral bins.  To keep computation time short only 10$^4$ photons per wavelength are simulated.
\begin{figure}
	\centering
	\subfigure[]{\includegraphics[width=5cm]{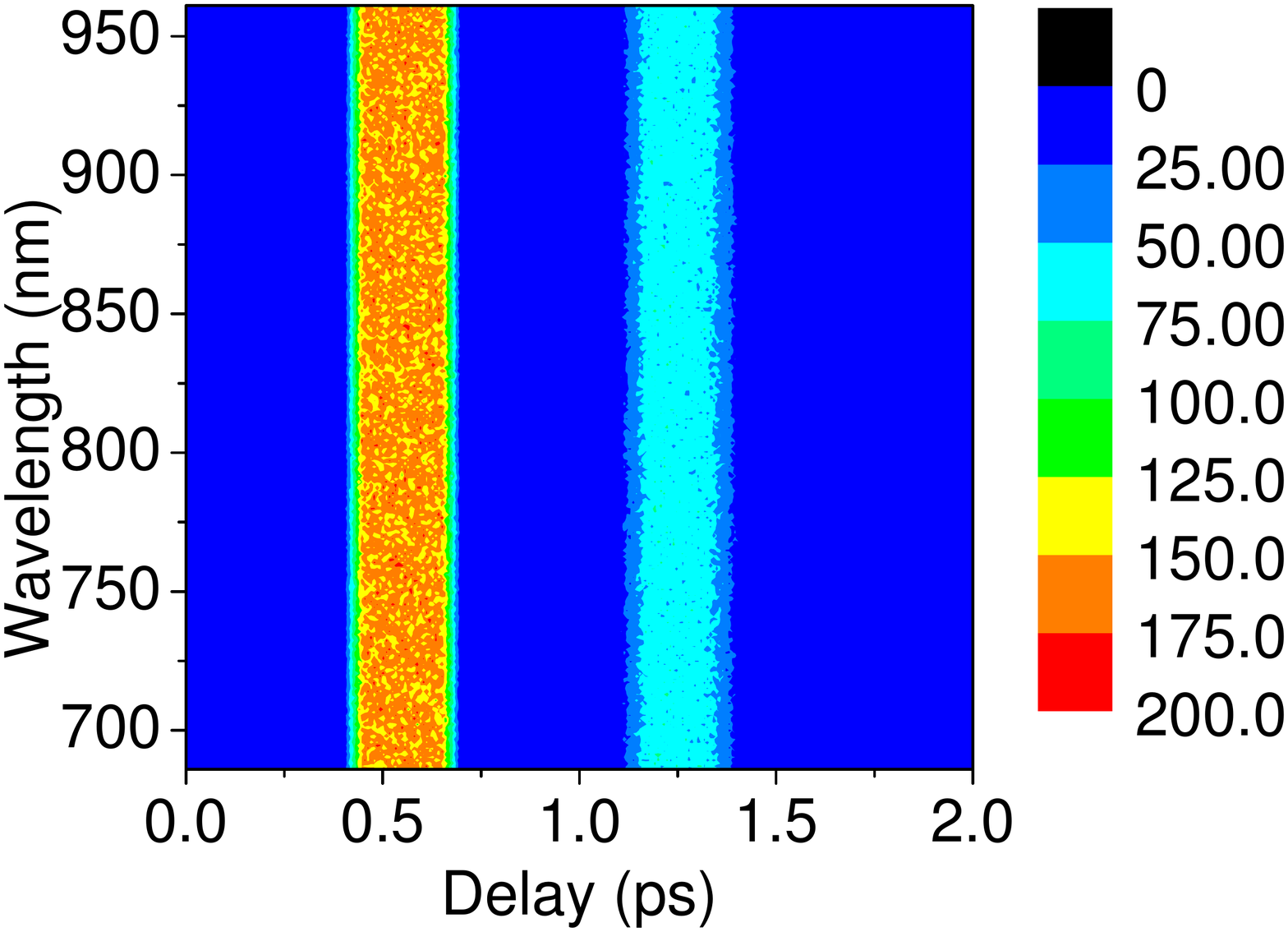}}
	\subfigure[]{\includegraphics[width=5cm]{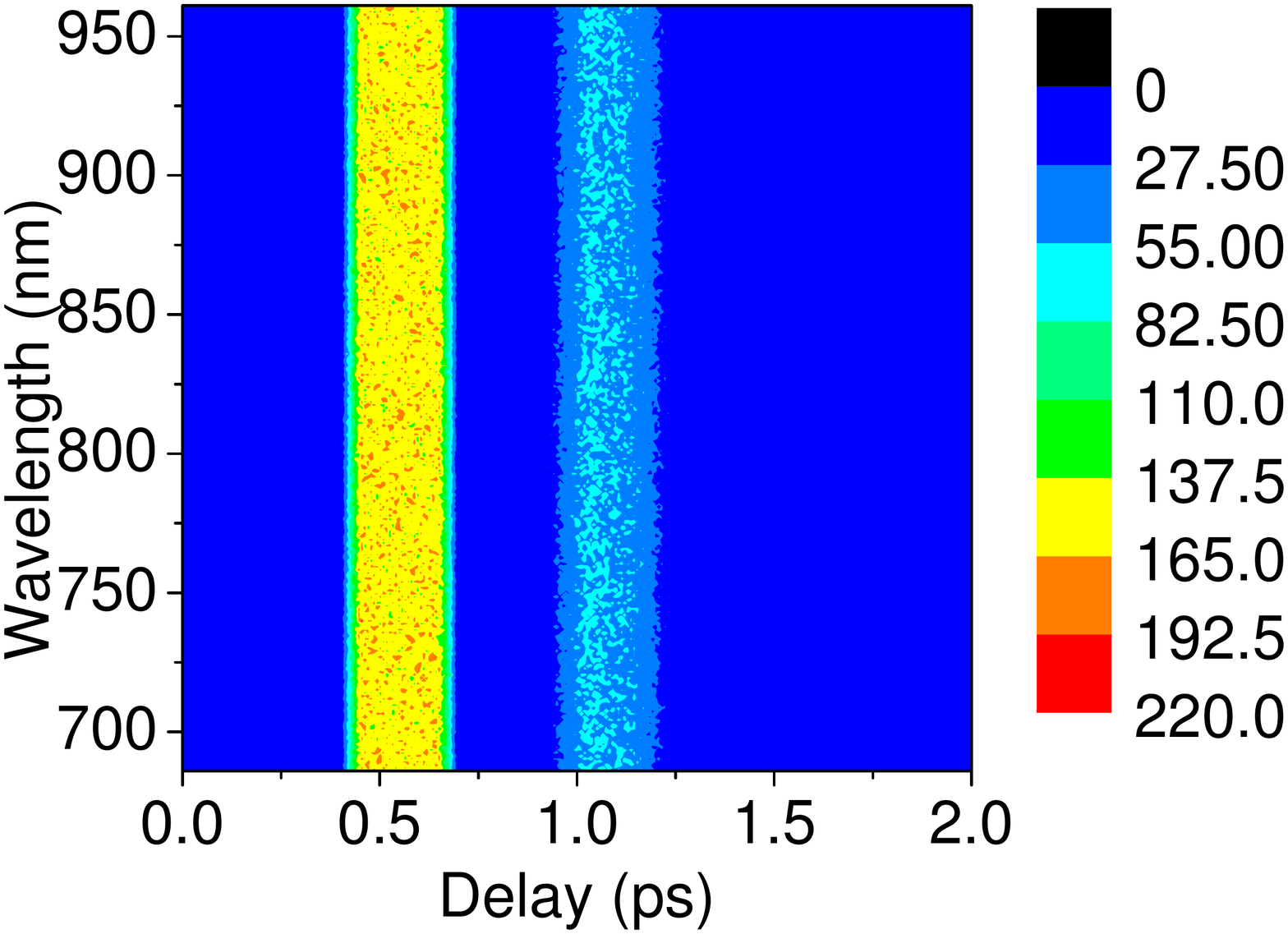}}\\
	\subfigure[]{\includegraphics[width=5cm]{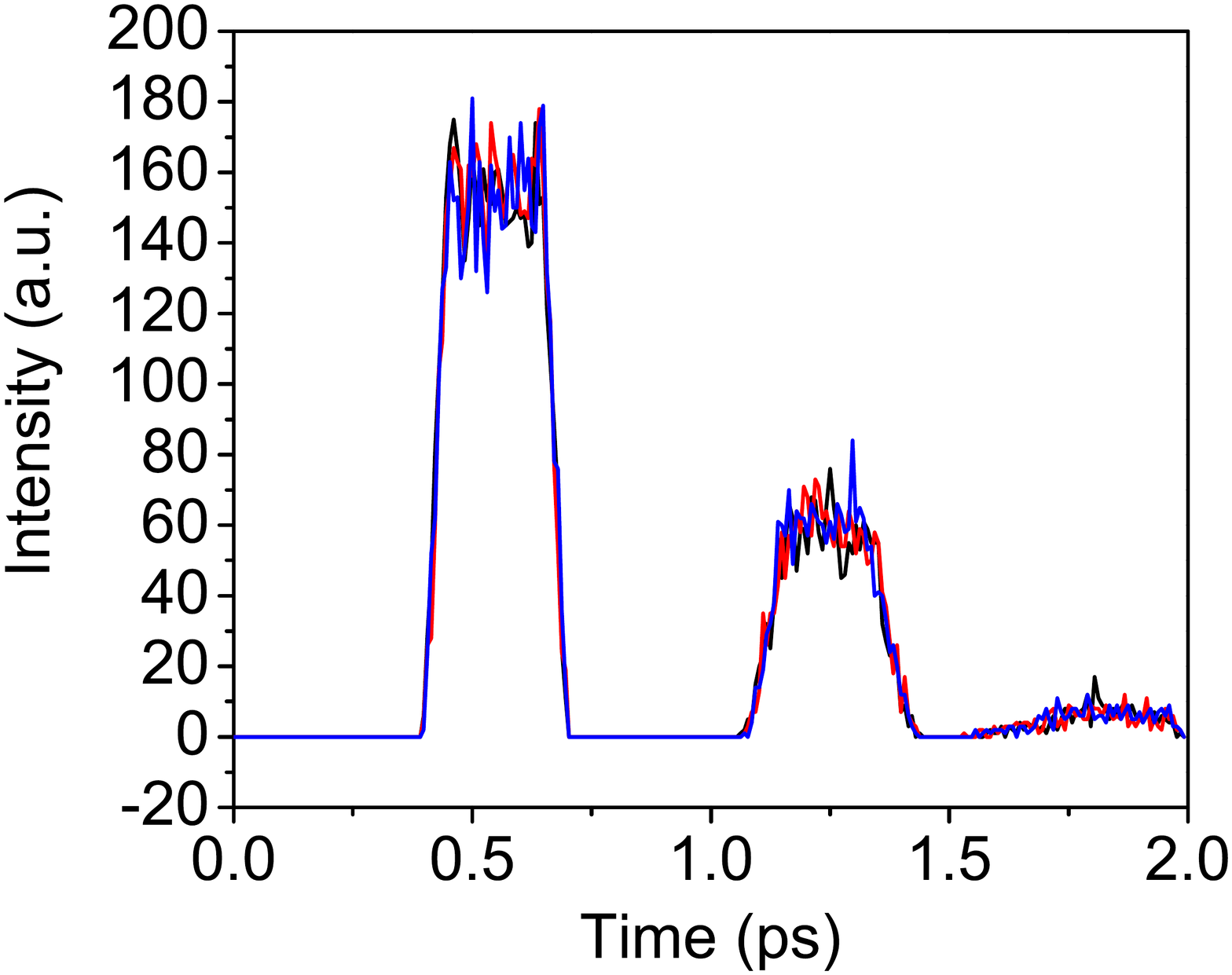}}
	\subfigure[]{\includegraphics[width=5cm]{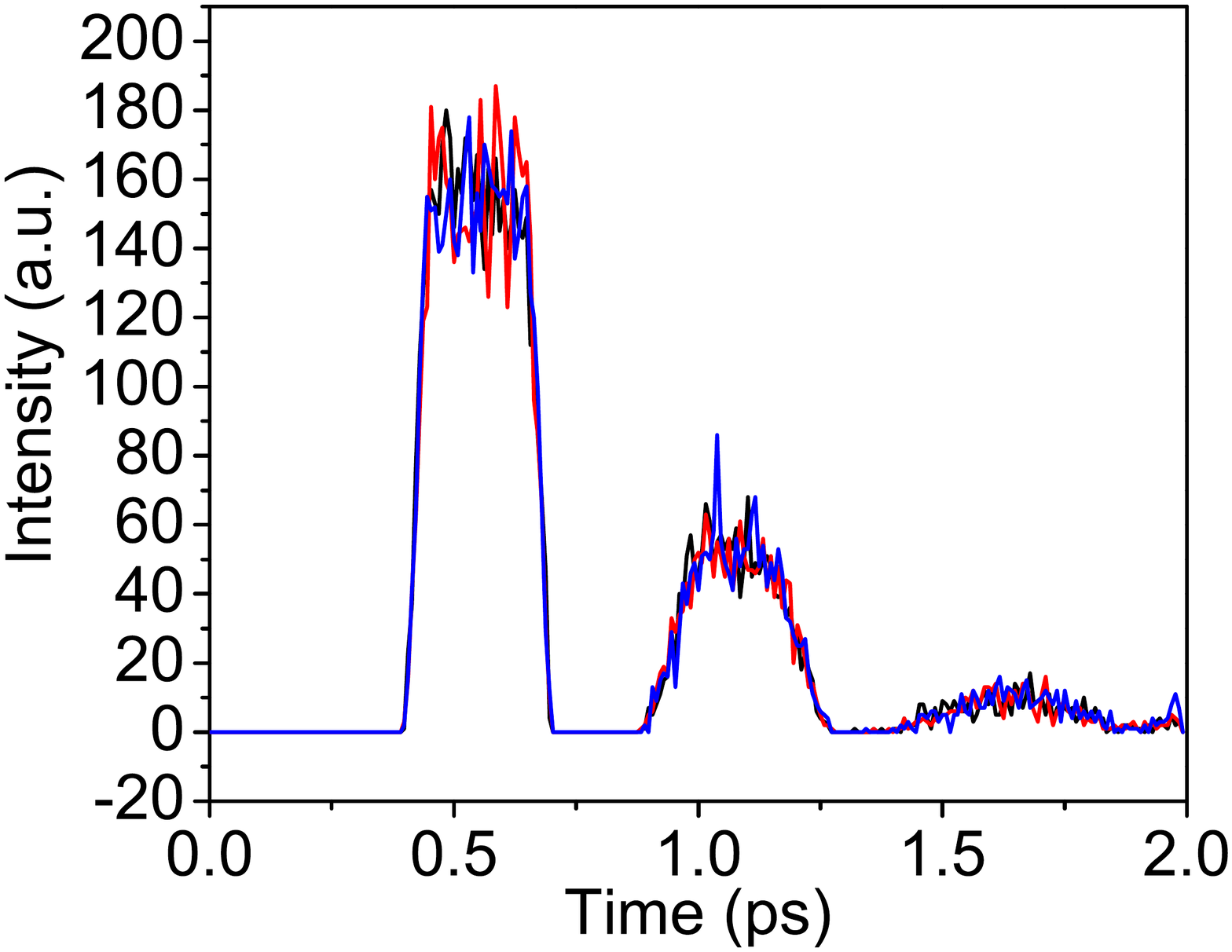}}\\
	\subfigure[]{\includegraphics[width=5cm]{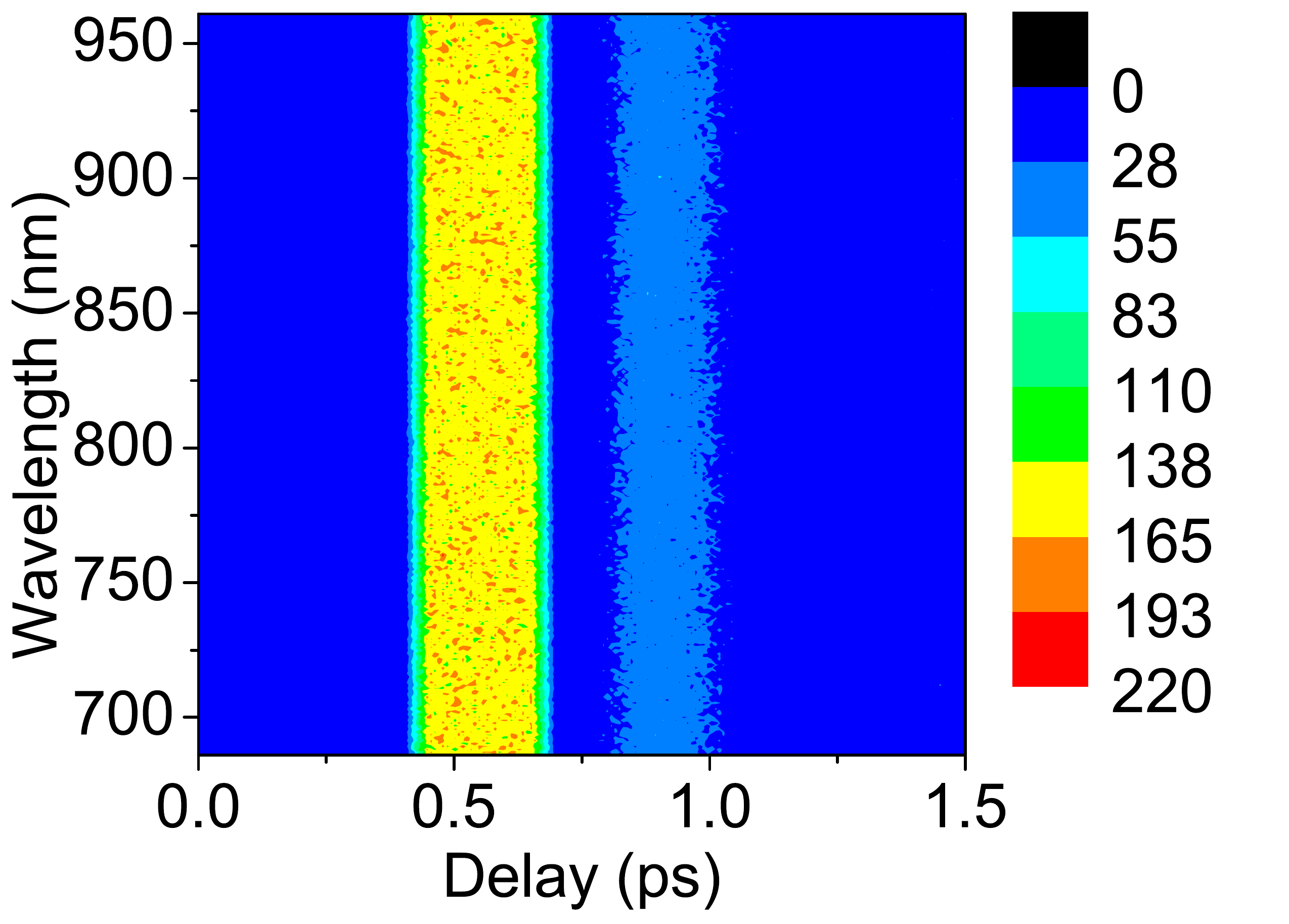}}
	\subfigure[]{\includegraphics[width=5cm]{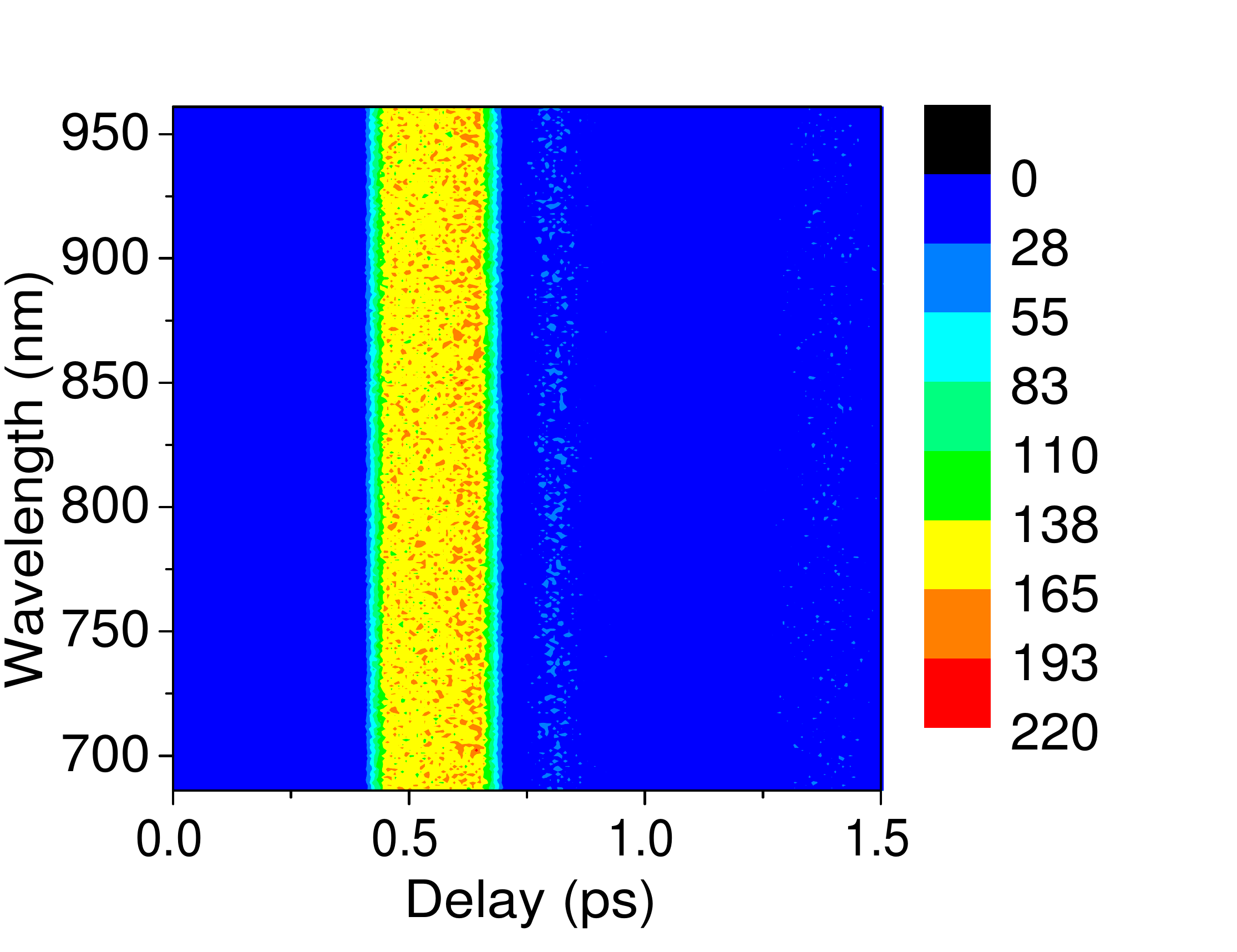}}
	\caption{The intensity as a function of frequency and time delay in a dispersive medium for 0.125 (a), 0.25 (b), 0.5 (d) and 0.75 (e) packing fractions. For comparison with Fig.~\ref{fig:Pulsedynamics}, line plots for 686 nm (black), 800 nm (red), and 961 nm (blue) for packing fractions 0.125 (c) and 0.25 (d) are also shown.  The particle system was 110~$\pm$~1~$\mu$m and the input pulse duration 50~fs for all simulations.}
	\label{fig:FROG1}
\end{figure} 

Figs.~\ref{fig:FROG1}(a)-(d) show comparison plots of the photon count as a function of wavelength and time delay for 110~$\pm$~1~$\mu$m particles at different packing fractions.  It is clearly seen that the delay time between the input pulse and first output pulse changes only slightly with wavelength and not at all by packing fraction.  However, the secondary pulses are quite drastically changed by the packing fraction, while showing little dependence on dispersion.  Although the secondary peaks are also considerably weaker than that shown in Fig.~\ref{fig:Pulsedynamics}, the relative strength between the main and secondary peaks is the same.  By varying the photon number, it was found that the ratio between the two peaks remains constant to within 5\%.  

\begin{figure}
	\centering
	\subfigure[]{\includegraphics[width=5cm]{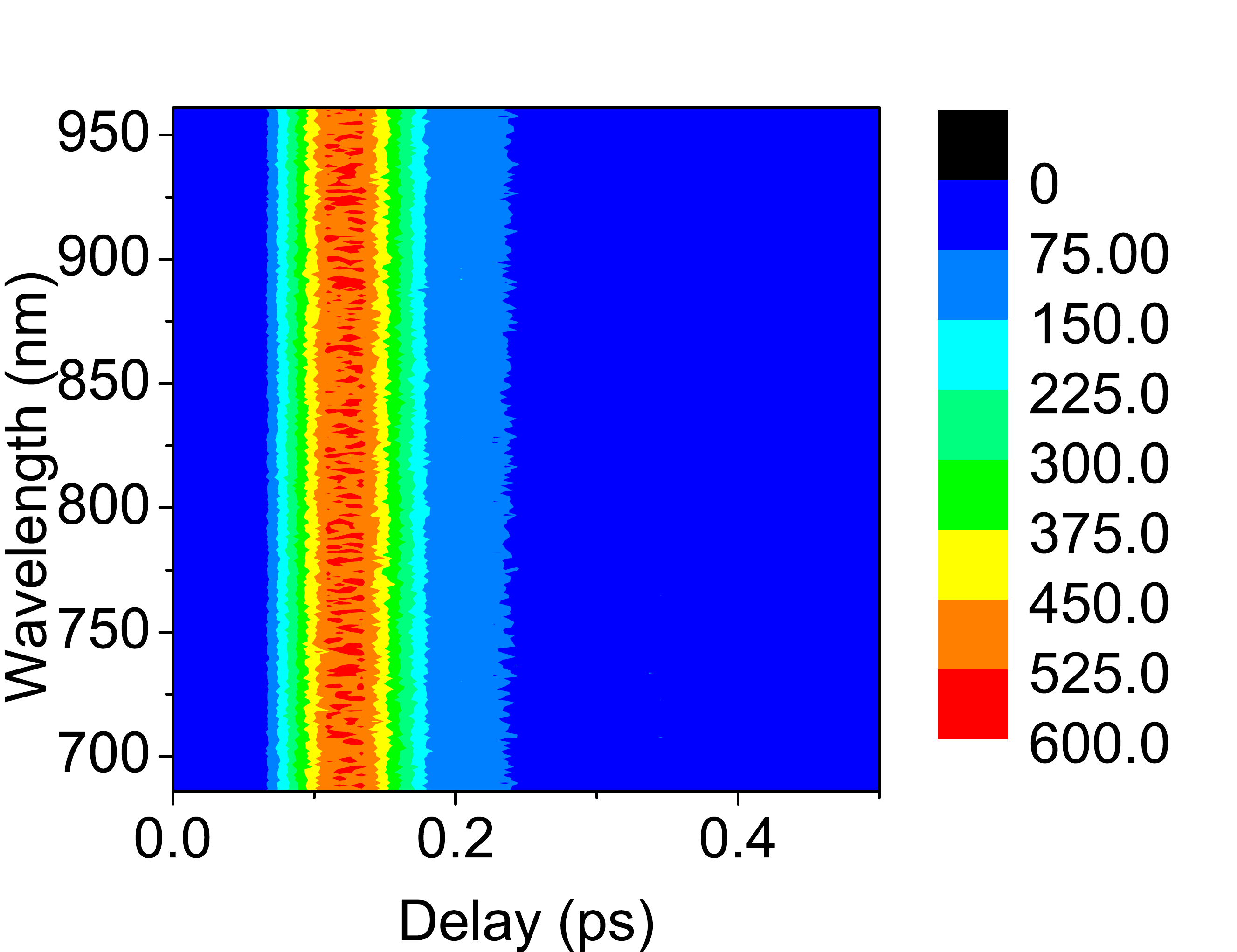}}
	\subfigure[]{\includegraphics[width=5cm]{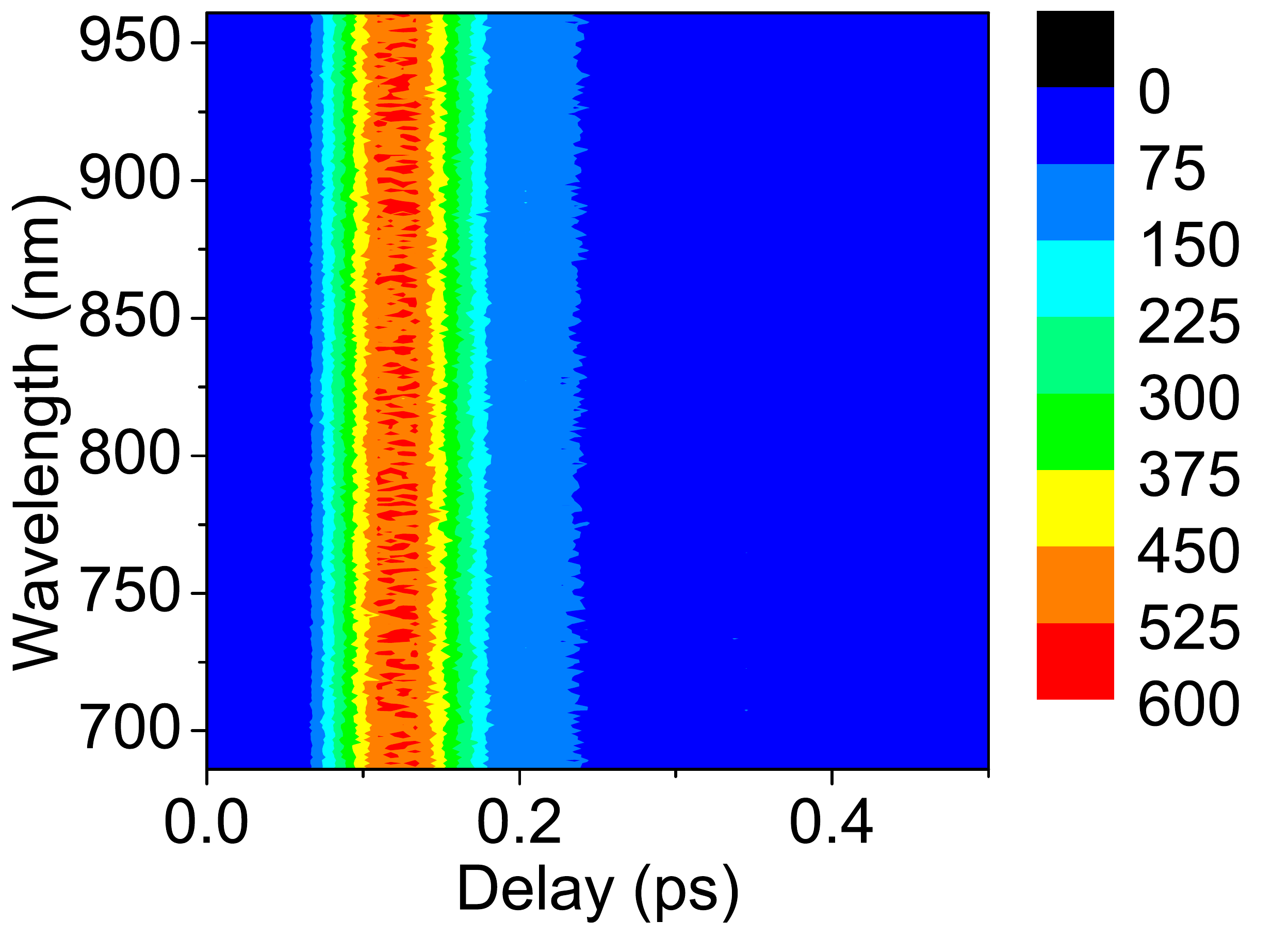}}\\
	\subfigure[]{\includegraphics[width=5cm]{frog05.pdf}}
	\subfigure[]{\includegraphics[width=5cm]{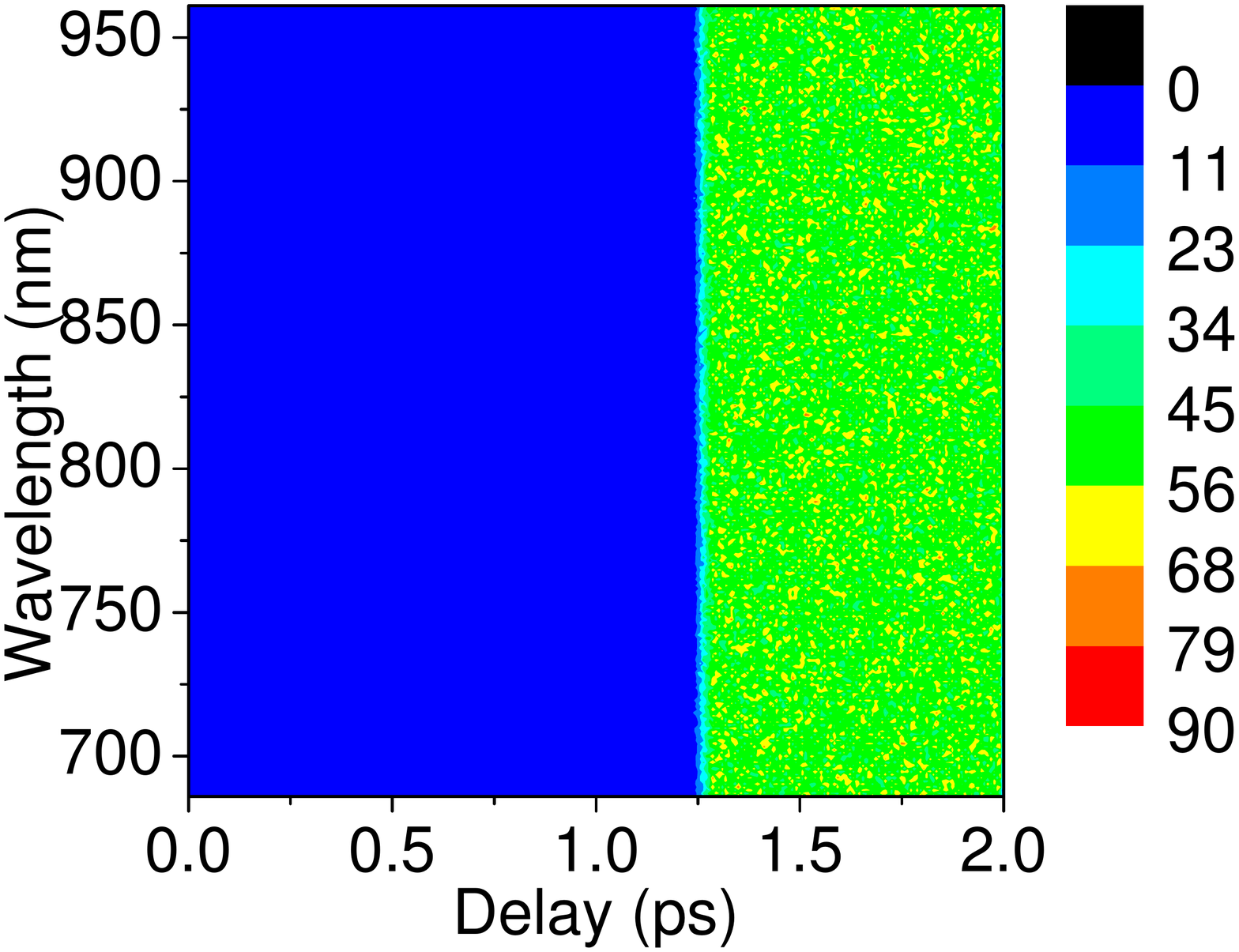}}
	\caption{The intensity as a function of frequency and time delay in a dispersive medium for 2.5~$\pm$~1~(a), 25~$\pm$~7.5~(b), 110~$\pm$~25~(c) and 325~$\pm$~70~(d)~$\mu$m particles.  The packing fraction was 0.5 and the input pulse duration 50~fs for all simulations.}
	\label{fig:FROG2}
\end{figure}

Figs.~\ref{fig:FROG2}(a)-(d) show frequency and time delay plots for 2.5~$\pm$~1, 25~$\pm$~7.5, 110~$\pm$~25 and 325~$\pm$~70~$\mu$m particle systems at a packing fraction of 0.5.  The change in delay time can be clearly seen from these plots, as can the pulse spreading due to changes in particle size ranges.  It is clear that the dispersion does not significantly alter the results found in subsection~\ref{ssec:no_dispersion}.  The time delay only changes slightly due to the wavelength dependent optical thickness of the scatterers.  Larger particles exhibit a slightly larger change in time delay with frequency simply because the photons must travel further between scattering events.      

\subsection{Dispersive, absorbing medium}\label{ssec:abs}
It is also of interest to see what effect the absorption of a medium may have on the time dependent scattering of light in the medium.  Here we are interested in the changes to the temporal shape of the backscattered pulse due the sudden change in dispersion associated with an absorption line.  Although an absorption line can be expected to change the temporal shape due to Beer's law, the influence of dispersion should be clearer if this is removed.  To investigate this, the dispersion of crown glass is again used and a further change in refractive index due to a broad (4000~GHz) Lorentzian line centered on 800~nm is added.  

\begin{figure}
	\centering
	\subfigure[]{\includegraphics[width=5cm]{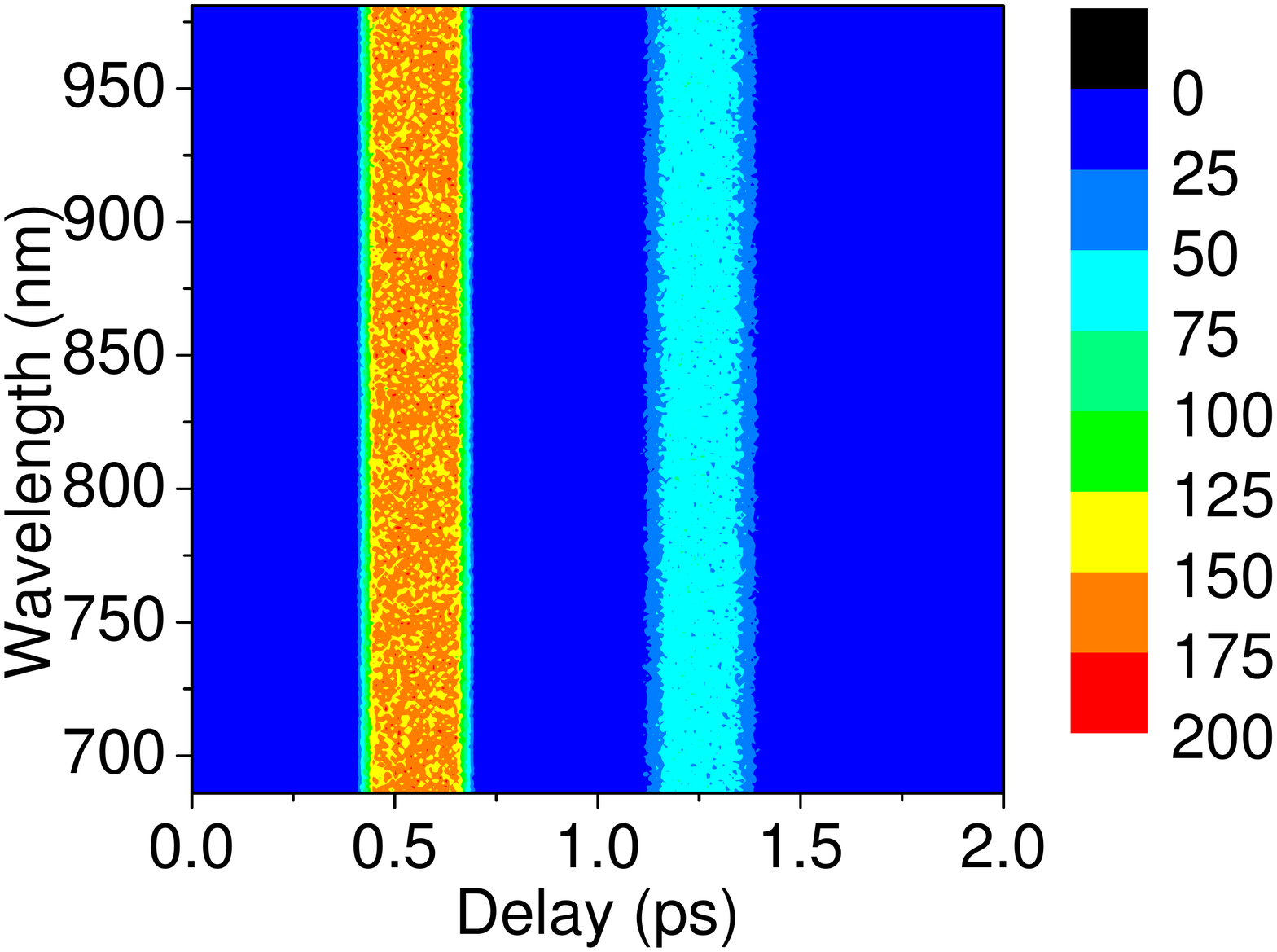}}
	\subfigure[]{\includegraphics[width=5cm]{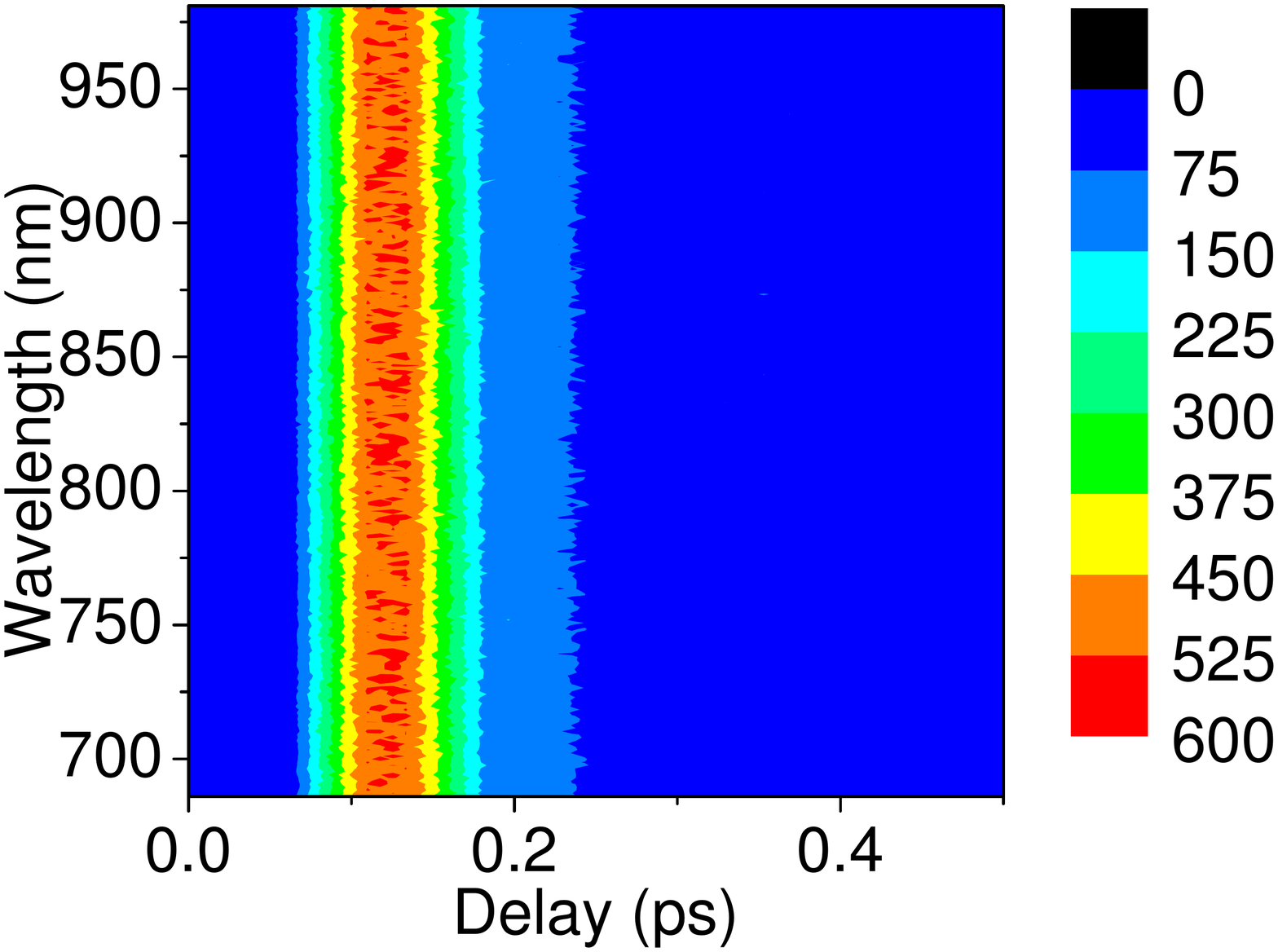}}
	\caption{The intensity as a function of frequency and time delay in a dispersive, absorbing medium.  110~$\mu$m,  0.125 packing fraction (a) and 25~$\pm$~7.5~$\mu$m, 0.5 packing fraction (b).  The input pulse duration 50~fs for both simulations.}
	\label{fig:FROG3}
\end{figure}

The refractive index changes as the derivative of the absorption feature, thus, causing an additional frequency dependent time delay.  However, as Fig.~\ref{fig:FROG3} shows, the change in time delay as a function of frequency is not readily observable.  This is because the subtle effects of the absorption feature are masked by those of the particle size distribution.  From this it can be concluded that a frequency resolved time-delay plot will not be significantly influenced by the chemical properties of the powder system, but rather is dominated by geometric scattering considerations.  Near infrared spectroscopy on solid state particulate systems shows that chemical absorption features are observable through the absolute photon number \cite{Blanco:1998}.  From this, it is expected that photon absorption would become observable in our simulations, provided that the frequency resolution is chosen to be sufficiently fine and photon absorption is included.  However, in our simulations, the frequency resolution has been taken to be typical for a frequency-resolved gating setup (500~GHz), which is much broader than typical rotational and vibrational overtone modes observed in the near infrared.  Thus, we can conclude that this method of detecting ultra-short pulses is insensitive to near infrared overtone modes, especially when the effects of scattering are included.

\subsection{Influence of the free path distribution}\label{ssec:fpd}
\begin{figure}
	\centering
	\includegraphics[width=7cm]{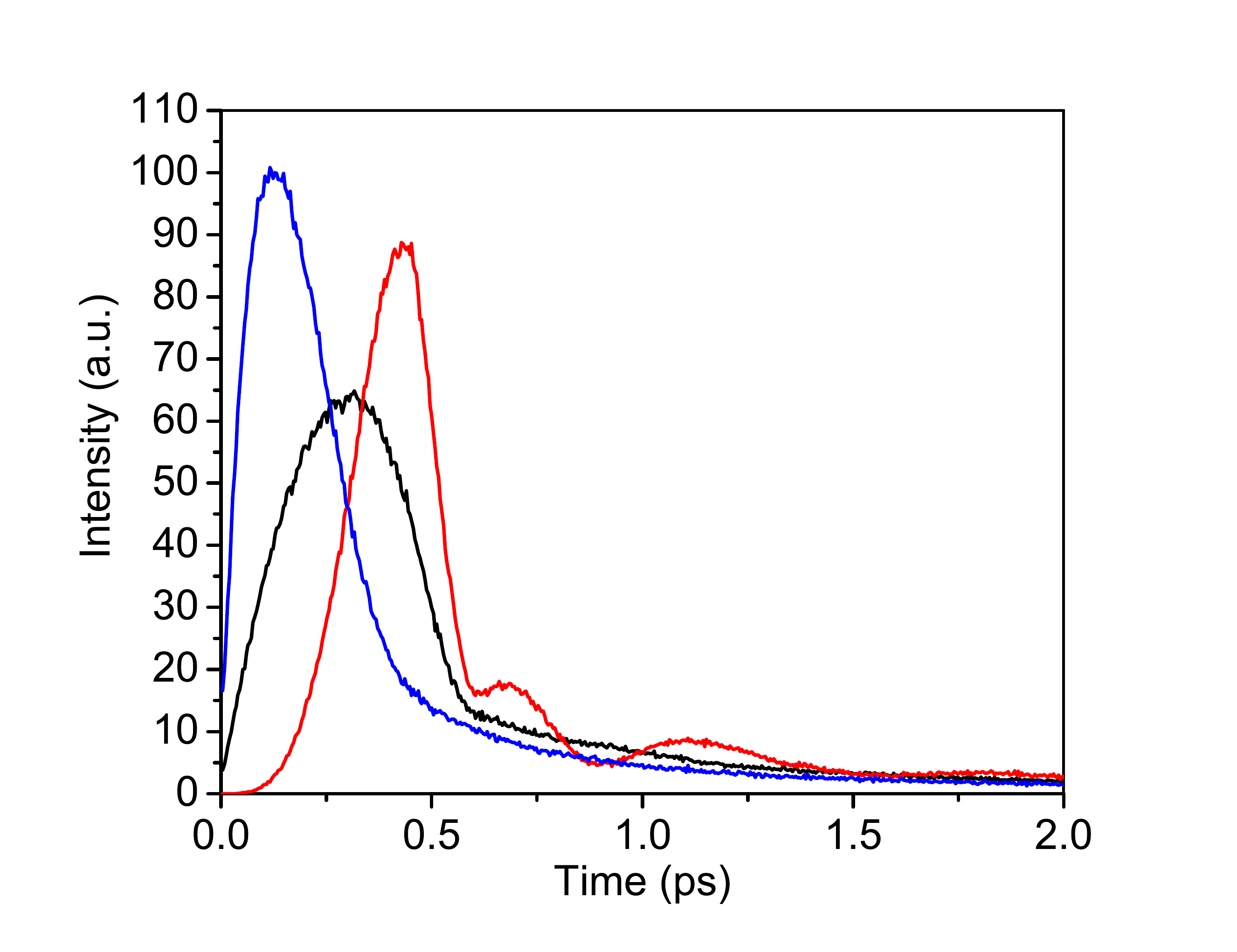}
	\caption{The exit pulse from a random media with a packing fraction of 0.5.  The particles are uniformly distributed between 85--135~$\mu$m.  The free path distribution is given by equation~\ref{eqn:beta}, with $\alpha = 2$, $\beta = 2$ (black), $\alpha = 2$, $\beta = 5$ (blue), and $\alpha = 5$, $\beta = 2$ (red).  }
\label{fig:fpOutputPulse} 
\end{figure}

As discussed in section~\ref{sec:Approach}, the free path distribution of the photons within the particles was originally approximated by the particle size distribution.  In this subsection the influence of the free path distribution is examined by assuming an uniform distribution (flat-top) particle size distribution, while the path length through a particle of size, $d$, is determined by equation~\ref{eqn:beta}.  From subsection~\ref{ssec:no_dispersion} we expect that the pulse shape of the scattered light will still be a convolution of the free path distribution and the input light pulse.  Figure~\ref{fig:fpOutputPulse} shows the typical output pulse from a medium consisting of scatters with an uniform size distribution and various free path distributions.  The particle size distribution can only be determined by the values for $\alpha$ and $\beta$.  It can be seen that the upper limit to the size range can still be observed as a discontinuity in the slope of the pulse, which combined with the slope of the rising and trailing edges can be used to constrain the particle size range.  It is also clear from Fig.~\ref{fig:fpOutputPulse} that time of flight measurements of the average particle size and packing fraction will be dominated by the free path distribution within the particles.

\begin{figure}
	\centering
	\subfigure[]{\includegraphics[width=5cm]{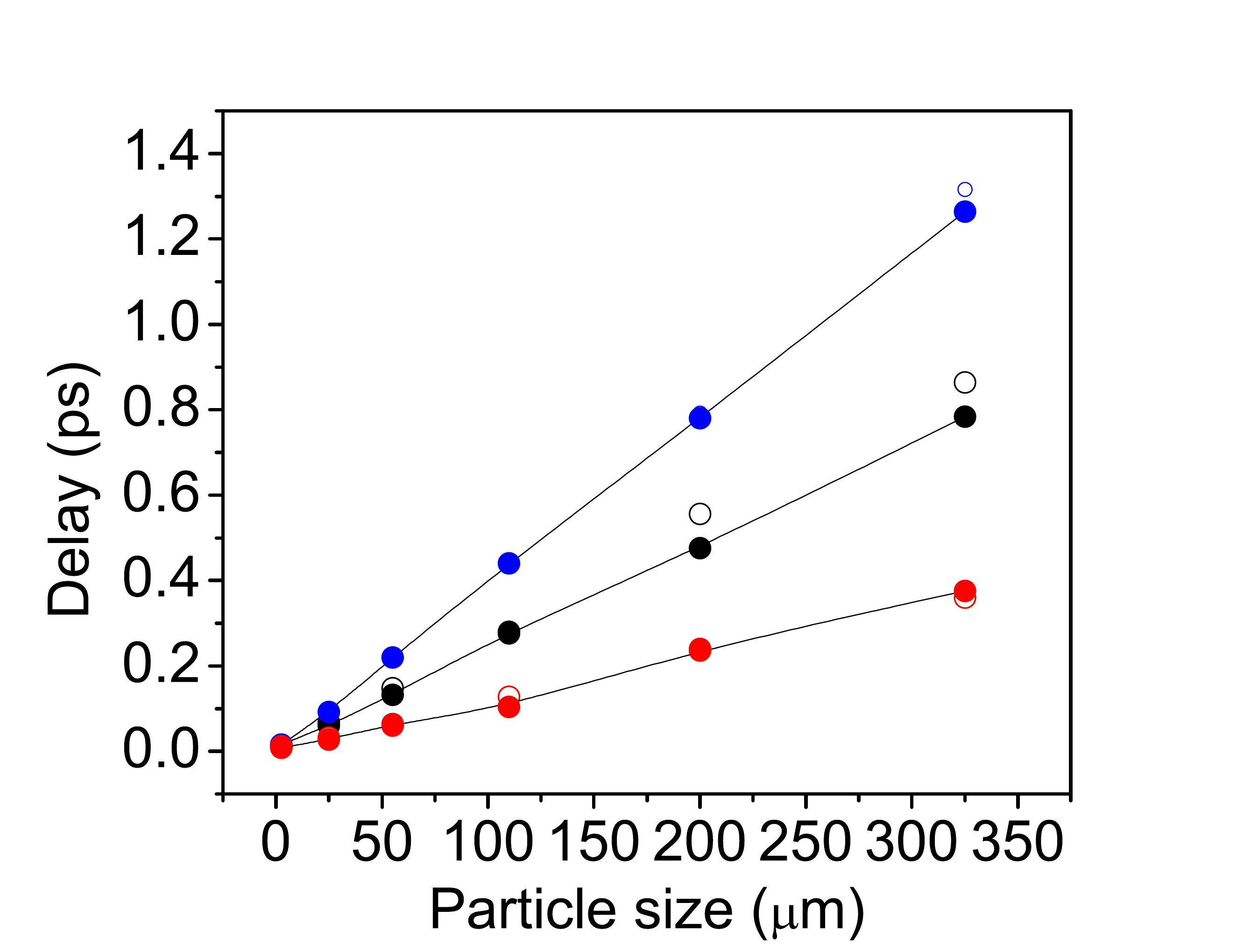}}
	\subfigure[]{\includegraphics[width=5cm]{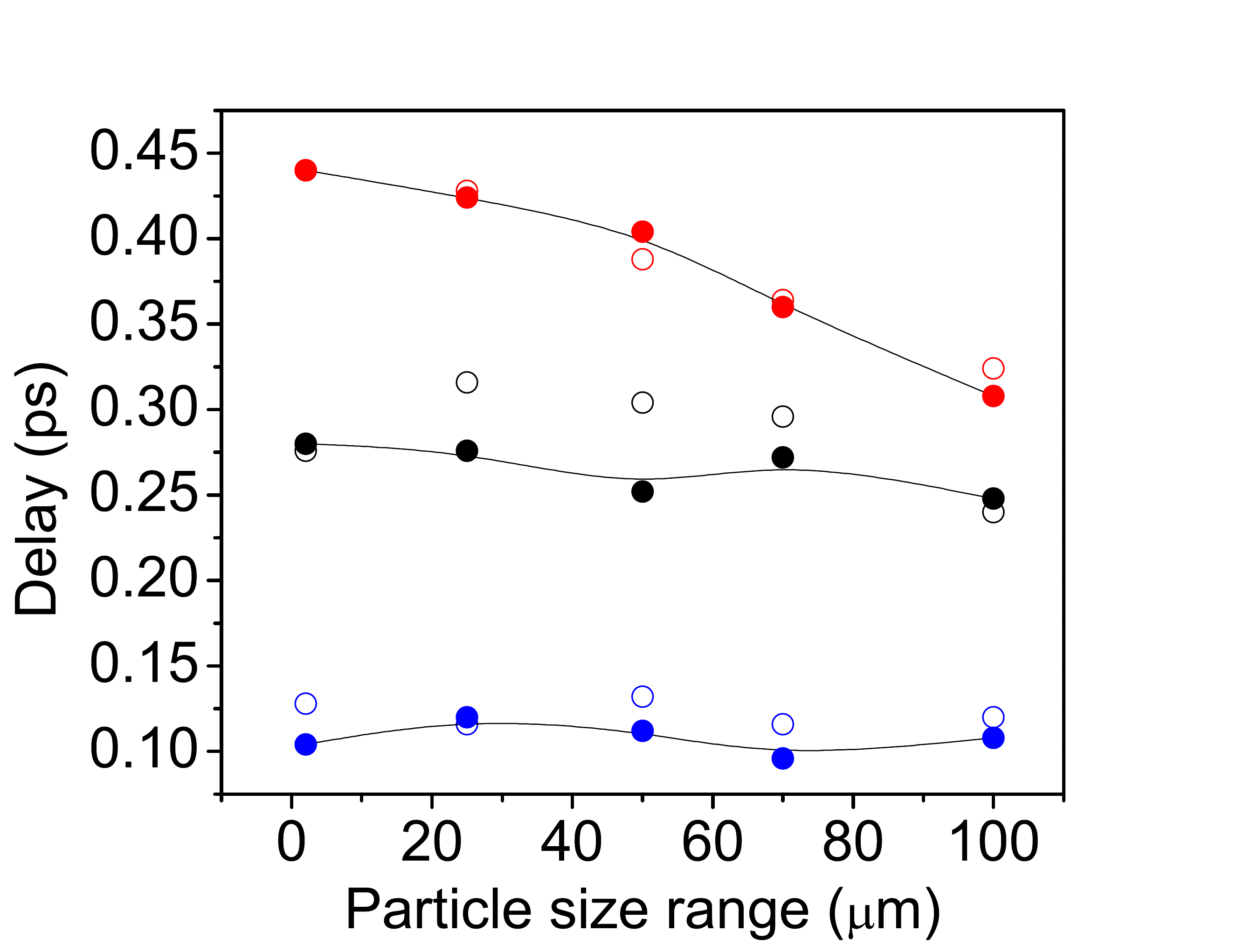}}
	\caption{The time delay as a function of particle size (a) and particle size range (b).  Open circles are for a packing fraction of 0.5, while closed circles are for a packing fraction of 0.9.  Three different free path distributions were used: $\alpha =2$,~$\beta = 2$ (black), $\alpha = 5$, $\beta = 2$ (red), and $\alpha = 2$, $\beta = 5$ (blue).  The lines are to guide the eye.}
	\label{fig:fpDelayTime}
\end{figure}

Despite these limitations, both quantitative and qualitative information can be still be obtained from the scattered light.  In Fig.~\ref{fig:fpDelayTime}(a), the time delay between the input pulse and the peak of the main pulse as a function of particle size is shown for three different free path distributions and two packing fractions.  The delay time is still independent of the packing fraction, however, it is clear that the time of flight measurement alone is unable to distinguish long needle-like particles ($\beta > 2$) from small spherical or cubic particles ($\alpha > 2$).  Figure~\ref{fig:fpDelayTime}(b) shows that the delay time is only very weakly dependent on the particle size range.  To obtain the average particle size, the influence of the free path distribution must be taken into account.  

\begin{figure}
	\centering
	\subfigure[]{\includegraphics[width=5cm]{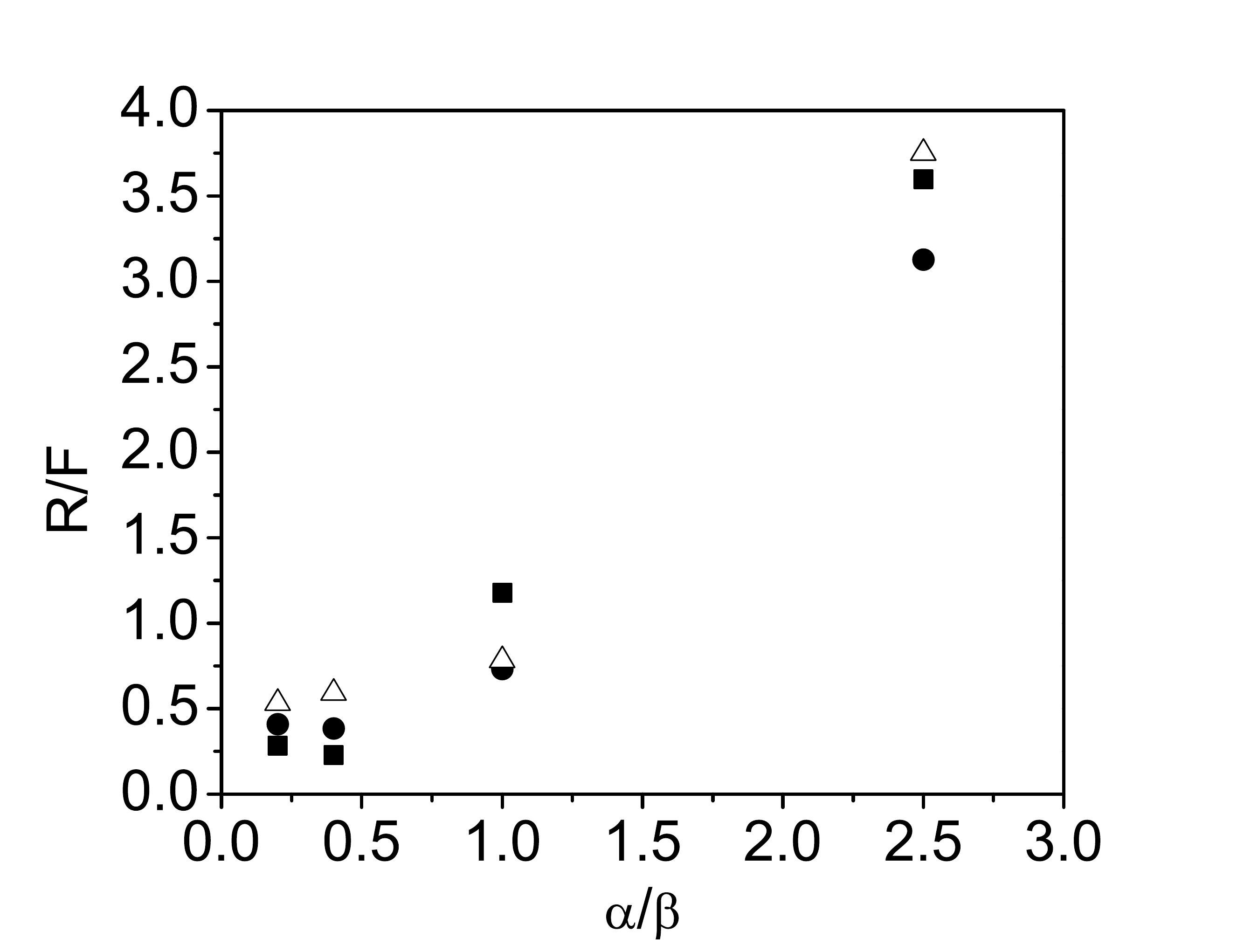}}
	\subfigure[]{\includegraphics[width=5cm]{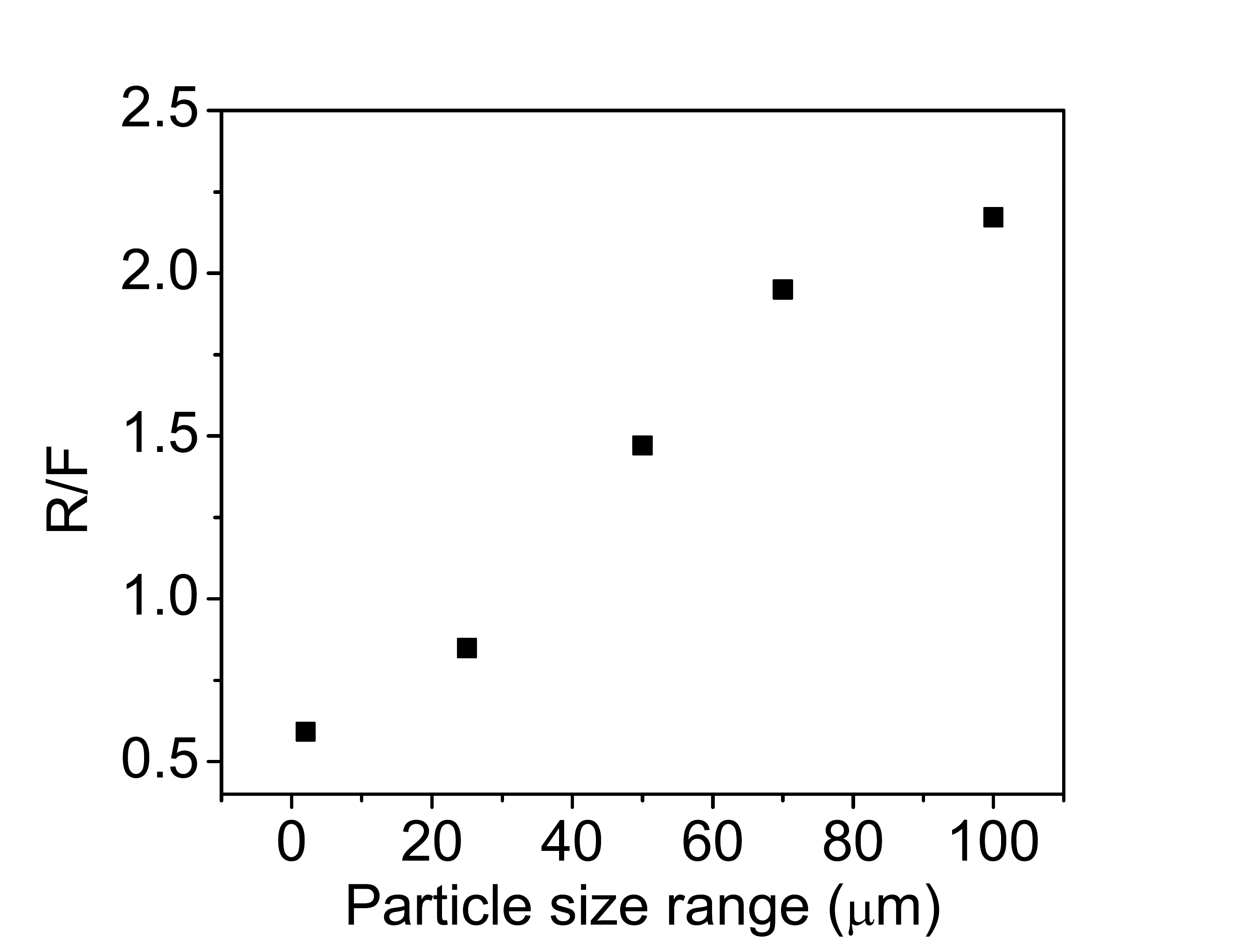}}
	\caption{The ratio of the rising and falling edge slopes (R/F) as a function of the ratio of $\alpha$ and $\beta$ (a) for 110~$\mu$m (triangles), 200~$\mu$m (circles), and 325~$\mu$m (squares) particles.  The size range was 2~$\mu$m.  The ratio of the rising and falling edge slopes as a function of particle size range (b) for $\alpha = 5$, $\beta = 2$, and an average particle size of 110~$\mu$m.}
	\label{fig:alphaBetaRatio}
\end{figure}

From Fig.~\ref{fig:fpOutputPulse} it is clear that the slope of the rising and falling edge of the main pulse also depends on the free path distribution.  This is illustrated in Fig.~\ref{fig:alphaBetaRatio}(a), where it can be seen that the ratio of the rising and falling edge slopes of the main peak depend on the shape of the free path distribution but not on the average particle size.  However, the slope also depends on the range of the particle size distribution (see Fig.~\ref{fig:alphaBetaRatio}(b)), which means that the particle size range must be obtained independently of the average particle size.

\begin{figure}
	\centering
	\subfigure[]{\includegraphics[width=5cm]{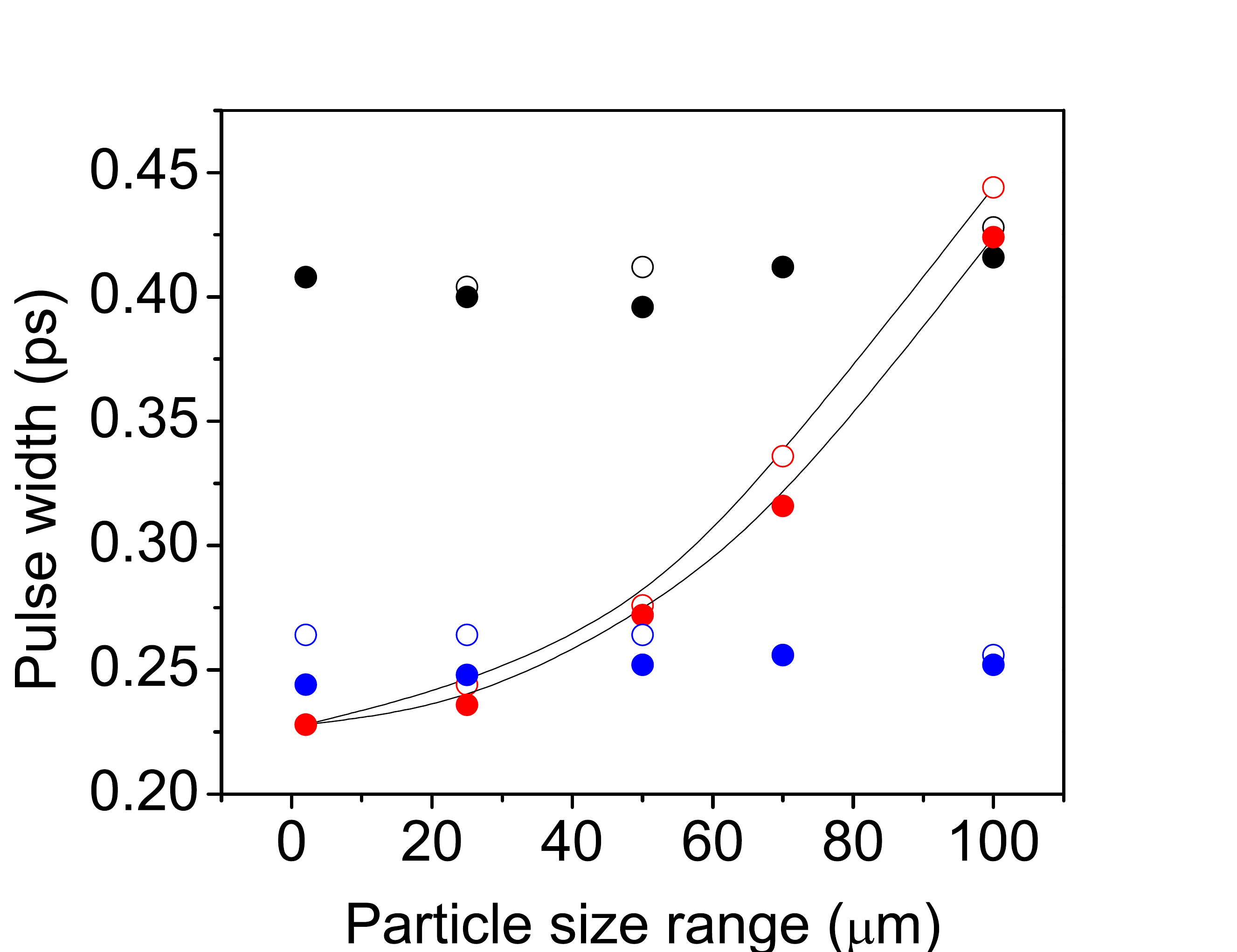}}
	\subfigure[]{\includegraphics[width=5cm]{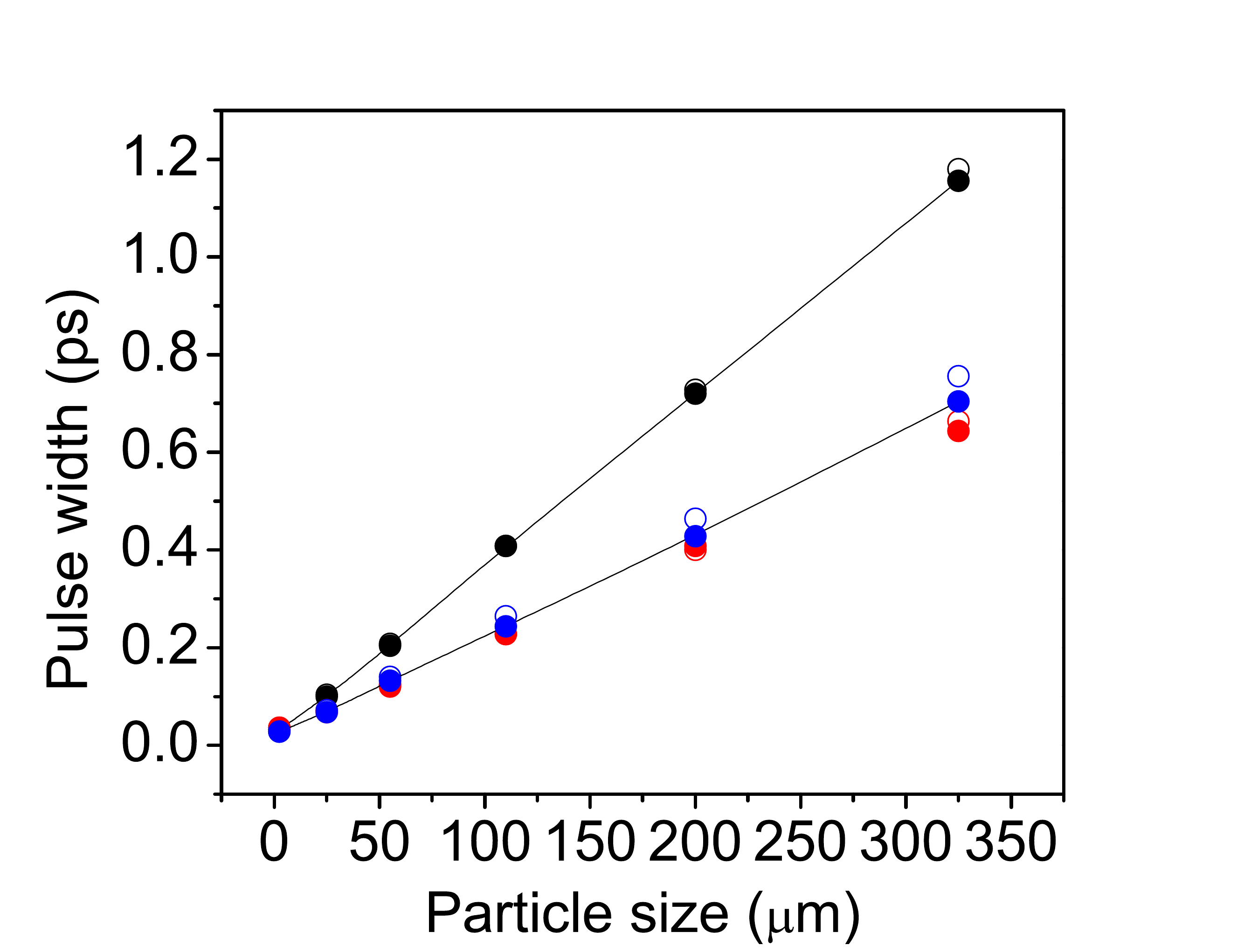}}
	\caption{The dependence of the pulse width on the particle size range (a) and the average particle size (b) for packing fractions of 0.5 (open circles) and 0.9 (closed circles).  The free path distribution is given by $\alpha = 2$,~$\beta = 2$ (black), $\alpha = 5$,~$\beta = 2$ (red), and $\alpha = 2$,~$\beta = 5$ (blue).  The lines are to guide the eye}
	\label{fig:pulseWidth}
\end{figure}

In subsection~\ref{ssec:no_dispersion}, the width of the particle size range was directly reflected in the width of the scattered pulse.  However, with the free path distribution the response is not so clear, Fig.~\ref{fig:pulseWidth}(a) shows that the dependence of the main pulse width on the particle size range is weak compared to the dependence on average particle size (Fig.~\ref{fig:pulseWidth}(b)).  In fact, in this situation, the pulse width represents a better way to measure the average particle size since it shows less dependence on the free path distribution.  The exception, $\alpha$,~$\beta = 2$, is one that does not correspond to most natural shapes, such as needle, cuboid, or spherical.  Even should such a free path distribution occur, it can be readily distinguished by the near unity value of the ratio of the rising and falling edge slopes.  Thus, the average particle size can be uniquely determined, which can then be used to obtain the particle size distribution from the ratio of the rising and falling edge slopes.

\begin{figure}
 	\centering
 	\includegraphics[width=7cm]{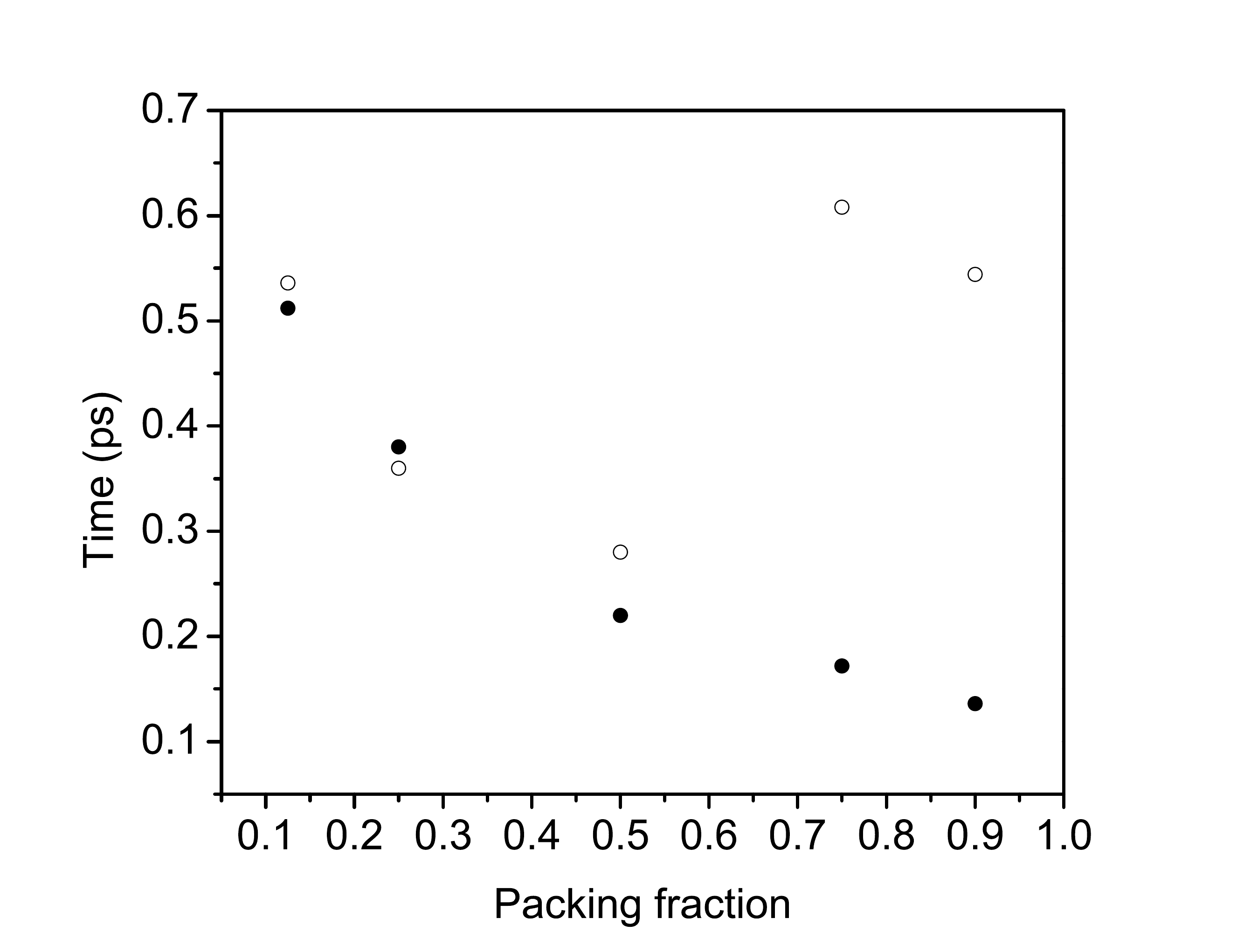}
 	\caption{The time delay between the first and second peaks of the scattered intensity for 110~$\mu$m~$\pm 1$~$\mu$m (closed circles) and $\pm 25$~$\mu$m (open circles).  The free path distribution parameters are $\alpha = 5$ and $\beta = 2$.}
 	\label{fig:fpPF}
 \end{figure}

 The delay between the initial peak and subsequent peaks is still dependent on the packing fraction (see Fig.~\ref{fig:fpPF}).  However, the free path distribution broadens the initial peak, causing the first and second peak to overlap for large packing fractions.  This can be seen by the increase in delay between first and second peaks at packing fractions of 0.75 and 0.9 for 110~$\pm$~25~$\mu$m particles.

\section{Conclusions}\label{sec:conc}
So far we have shown that the backscattering of near-infrared ultra-short pulses by random media looks rather promising for rapid (on-line) monitoring, e.g., of aggregation during a granulation process. In particular, the specific temporal shape of the backscattered pulses reveals information on the average particle size distribution and the packing density. This raises the question of what detection techniques would record the shape of the returning pulses with a sufficiently high speed for on-line monitoring.  The only methods to measure the temporal shape of ultra-short pulses on a fs time scale are optically nonlinear techniques. However, in the case considered here there are some fundamental problems to overcome before these methods can be employed in a straightforward manner. Usually these detection methods only work well when analyzing light pulses that propagate in a single spatial mode and with sufficient power. In the present situation, however, the power of the light is rather low because it originates from scattering (at most a few percent of the incident light). More importantly, the light to be temporally resolved is distributed over a large number of spatial modes due to a large source area (several mm$^2$) and a wide ($2\pi$) solid angle. Furthermore, the temporal shape in each spatial mode would be different from that of other modes because it only contains the time-of-flight contributions from an individual set of scattering paths, and can thus substantially deviate from the average pulse shape generated by the sample. To obtain the representative (path averaged) pulse shape, one requires a measurement that integrates over all spatial modes of the scattered light. 

The required temporal resolution within a single backscattered mode can be obtained by time-resolved optical gating with an ultra-short reference (gating) pulse \cite{Baigar:1998}. Typically in such gating a narrow solid angle of scattered light is collected and passed through a nonlinear crystal along with an ultra-short reference (gating) pulse, where the second harmonic is generated. The average power of the second harmonic versus the delay of the reference pulse is recorded with a slow detector. This technique has been shown to be successful for imaging objects embedded within strongly scattering media \cite{Baigar:1998}.  However, the small acceptance angle of phasematching and the low power of the scattered light leads to recording times of minutes for a single spatial mode, and this translates into multiple hours of recording when the average temporal shape is recorded. As a result, standard nonlinear techniques based on optical gating in an external nonlinear crystal are not well suited for real-time pulse-shape monitoring in industrial processes.

We have presented a simple model to describe the backscattering of ultra-short pulses in random media. The simulations predict that the most important physical characteristics of the scattering medium, i.e., its particle size distribution and packing density, can be retrieved by recording the temporal shape of the backscattered pulse. We note that this is in qualitative agreement with our previous experimental results \cite{Hauger:1997}. Specifically, for a single injected ultra-short pulse we predict that the response consists of multiple pulses. The shape of the main returning pulse reflects that of a convolution between the free path distribution within the particles and the particle size distribution function. The FWHM pulse width of the main output pulse depends on the particle size, while the ratio of the rising and falling edge slopes can be used to determine the particle size range. The separation between the main peak and the secondary peak depends on both the packing fraction and the particle size distribution. With these properties, the output response contains enough information to obtain the average particle size, the particle size distribution and packing fraction of the scatterers, without knowledge of the shape of the particles. These results are not significantly influenced by normal dispersion or spectrally dependent refractive index changes due to absorption features. 

Our investigations indicate that time dependent scattering of ultra-short pulses has significant potential for rapid in-line characterization of particle size range and density, which can support the analysis of spectroscopic techniques.
\end{document}